\documentclass[aps,pra,preprint,superscriptaddress,longbibliography]{revtex4-1}

\usepackage{amsmath}
\usepackage{amsthm}
\usepackage{amsfonts}
\usepackage{amssymb}
\usepackage{xcolor,graphicx}
\usepackage{tikz}
\usepackage{color}
\usepackage[pdftex]{hyperref} %para ligar las referencias
\usepackage{longtable}
\usepackage{array}
\usepackage{dsfont}

\begin{document}
	
%--------------------------------------------------------------------------------
\title{Reconfigurable multiport switch in coupled mode devices}

\author{B. Jaramillo-\'Avila}
\email[e-mail: ]{jaramillo@inaoep.mx}
\affiliation{CONACYT-Instituto Nacional de Astrof\'{i}sica, \'{O}ptica y Electr\'{o}nica, Calle Luis Enrique Erro No. 1. Sta. Ma. Tonantzintla, Pue. C.P. 72840, Mexico}

\author{B. M. Rodr\'iguez-Lara}
\email[e-mail: ]{bmlara@tec.mx}
\affiliation{Tecnologico de Monterrey, Escuela de Ingenier\'ia y Ciencias, Ave. Eugenio Garza Sada 2501, Monterrey, N.L., Mexico, 64849 \\ Instituto Nacional de Astrof\'{\i}sica, \'Optica y Electr\'onica, Calle Luis Enrique Erro No. 1, Sta. Ma. Tonantzintla, Pue. CP 72840, Mexico}

\date{\today}

	\begin{abstract}
	We consider a coupled mode system where the effective propagation constants of localized modes are amenable to modulation.
	Starting from an unmodulated system where power transfer is heavily suppressed, we demonstrate that small, periodic modulation of the propagation constants enhance power transfer using a slowly varying envelope approximation for the field mode amplitudes.
	We calculate an approximate modulation frequency enabling complete transfer between otherwise negligibly coupled elements. 
	The ability to control these modulations by electrical or thermal effects allows for reconfigurable multiport switching.
	We use an array of coupled silica waveguides and the thermo-optic effect to test our predictions in the telecom C-band. 
	However, this requires a refractive index modulation with period of the order $~10^{-3}~\mathrm{m}$ 
	and yields total power transfer with a propagation distance of the order of $10^{-1}~\mathrm{m}$, 
	which might make it unattractive for integrated photonic applications.
	Nevertheless, our results are valid for devices described by an equivalent coupled mode matrix for space or time propagation; for example, arrays of microring or terahertz resonators, microwave cavities, radio frequency antennas, or RLC circuits.
	\end{abstract}

%\pacs{05.45.Mt, 42.50.Ct, 42.50.Mn, 73.43.Nq}

\maketitle
%--------------------------------------------------------------------------------
%%%%%%%%%%%%%%%%%%%%%%%%%% body %%%%%%%%%%%%%%%%%%%%%%%%%%
\section{Introduction}

Multiport switching is essential in both classical and quantum optics applications. 
It is useful to route optical signals in photonic devices and provides a technological platform for photon-based programmable quantum computers \cite{Matthews2009,Carolan2015,Wang2019}. 
Optical switch design relies on a plethora of physical systems, including micro-electro-mechanical \cite{Giles1999,Borovic2004,Yano2005}, complementary metal oxide semiconductor \cite{Tsybeskov2009,Rylyakov2012}, and evanescently coupled waveguides. 
We focus on the latter as it displays a strong similarity with finite dimensional quantum mechanical systems and allows us to exploit optical analogues of quantum mechanical effects \cite{Longhi2009} and abstract symmetries \cite{RodriguezLara2015b,Villanueva2015,RogriguezLara2018}. 

In waveguide arrays, switching may be induced by coupling strength modulation while keeping the refractive indices constant. 
This produces optical analogues, for example, of stimulated Raman adiabatic passage (STIRAP) \cite{Longhi2006,Longhi2007,Della2008}, where coupling strength modulation occurs along the propagation direction, or optical realizations of $\mathbf{J}_{x}$ Hamiltonians \cite{PerezLeija2013}, where the modulation is constant along the propagation direction but varies from waveguide to waveguide. 
Other approaches vary the effective refractive indices of individual waveguides while keeping the couplings constant. 
This produces optical analogues of Anderson localization of light \cite{John1984,DeRaedt1989,Schwartz2007,Segev2013,Jaramillo2019}, where the refractive index of individual waveguides varies randomly, or parity-time symmetry \cite{Ruschhaupt2005,ElGanainy2007,HuertaMorales2016,Nodal2018}, where refractive indices include gain or loss following a certain structure. 
Finally, other approaches vary both couplings and refractive indices \cite{RodriguezLara2014,RodriguezLara2014b,RodriguezWalton2019}.

Here, we are interested on the latter and propose a scheme where coupling strengths are constant but vary from waveguide to waveguide following an abstract symmetry plus periodical modulation of individual refractive indices.
Our symmetry-based proposal complements grating assisted couplers \cite{Marcuse1987,Huang1989,Griffel1991,Alferness1992,Weisen2019}, where dissimilar waveguides are coupled by resonant periodic variations in the effective refractive index. 
The coupling allows near complete power transfer between the waveguides. 
Without the grating, the waveguides would have negligible power transfer due to high effective refractive index differences between them.
These gratings can be fixed \cite{Marcuse1987,Huang1989} or reconfigurable \cite{Alferness1992}.
This may also be reminiscent of $\Delta \beta$ switches \cite{Kogelnik1975,Kogelnik1976,Schmidt1979,Korotky1986,Findakly1988}, 
where a switch designed for a given wavelength can work with a different one by adjusting an alternating constant phase shift induced by the electro-optic effect through a series of electrodes.

Our proposal starts from an array where power transfer is negligible due to high refractive index differences between individual waveguides. 
We use the underlying symmetry and harmonic modulation of refractive indices to induce power transfer in a manner analogous to population inversion due to optical driving in quantum systems \cite{Mandel1995}.
Controlled changes of the refractive index can be induced by electro- \cite{Turner1966,Long1994} or thermo-optic \cite{Gao2018} effects, for example. 
The former has applications in building modulators \cite{Liu2015}. 
The latter is widely used to produce controlled phase shifts in directional couplers in quantum photonic circuits \cite{Matthews2009,Carolan2015,Wang2019}. 
Electro- and thermo-optic modulation produce small refractive index changes of the order of $10^{-10}$ per volt/meter \cite{Turner1966,Long1994} and $10^{-5}$ per kelvin \cite{Gao2018}, in that order. 
For such small variations of the refractive index, power transfer, or switching, occurs over long propagation distances. 
The level of control and the long propagation distances required to produce reconfigurable multiport switching might render our proposal unfeasible for integrated photonic devices.
However, we hope that theoretical curiosity and its inherent first principles validity for systems described by coupled mode theory, either in space propagation or time evolution, are enough to motivate the exploration of such an avenue in, 
for example, microring \cite{Liu2005} or terahertz \cite{Preu2008} resonators, microwave cavities \cite{Haus1991,Elnaggar2014,Elnaggar2015}, radio frequency antenna \cite{Kim2007}, or RLC circuits \cite{Agarwal2006}. 

In the following, first, we present a theoretical model describing an array of waveguides with an underlying $SU(2)$ symmetry where different individual refractive indices strongly limit the power transfer produced by evanescent coupling. 
Then, we add small, controllable, periodic changes on the refractive index of the waveguides and show that the amplitude and frequency of these changes produce resonant power transfer between the waveguides. 
We show that it is possible to analytically approximate this power transfer using a slowly varying envelope approximation over long propagation distances. 
Next, we go beyond the coupled mode theory approximation and
use a 2D finite element simulation to demonstrate
power transfer between waveguides with our approach. 
Finally, we discuss the implications and requirements of our proposal.

%%%%%%%%%%%%%%%%%%%%%%%%%%%%%%%%%%%%%%%%%%

\section{Model}

Coupled mode theory provides a tractable framework to study light propagation along some spatial or temporal dimension $\zeta$ \cite{Snyder1972,McIntyre1973,Huang1994}; for example, evanescently coupled arrays of waveguides or ring microresonators, in that order. 
The coupled mode equation,
\begin{equation}
- i\, \frac{\mathrm{d} \vec{\mathcal{E}}(\zeta)}{\mathrm{d}\zeta} = \mathbf{M}(\zeta) \cdot \vec{\mathcal{E}}(\zeta),
\end{equation}
describes the dynamics of polarized modes localized within the $d$ elements of the array. 
The field in the $j$-th element is $\vec{E}_{j}(x,y,\zeta) = \hat{\epsilon}\, \psi(x,y)\, \mathcal{E}_{j}(\zeta)$.
There, $\hat{\epsilon}$ gives the polarization and $\psi(x,y)$ gives the spatial mode profile. 
The $d$-dimensional vector $\vec{\mathcal{E}}(\zeta)$ collects the modal amplitudes $\mathcal{E}_{j}(\zeta)$. 
The coupled mode matrix $\mathbf{M}(\zeta)$ diagonal and off-diagonal elements contain the effective propagation constants corresponding to each element-localized mode and the coupling between pairs of them, in that order. 
They may depend on the variable $\zeta$ or not.

\begin{figure}[ht!]
	\centering
	\includegraphics[scale=1.5]{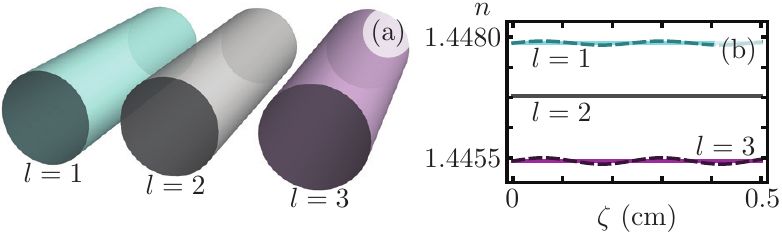}
	\caption{(a) Waveguide array that generates a three-dimensional coupled mode matrix $\mathbf{M}$ and (b) the corresponding refractive indices. The difference in propagation constants, given by $\left[ \Delta \beta + f(\zeta) \right] \mathbf{J}_{z}$, is generated by the different refractive indices in (b). The solid lines yield unmodulated propagation constants, $f(\zeta)=0$, while the dashed curves yield a sinusoidal modulation, $f(\zeta)=\delta \sin(\omega \zeta)$. The off-diagonal couplings $2g\, \mathbf{J}_{x}$ are controlled by the separation between the waveguides. The difference between off-diagonal coupling constants in the modulated and unmodulated cases is very small and therefore neglected.}\label{fig:1}
\end{figure}

In order to provide multiport switching, we choose an array given in terms of a finite dimensional representation of the angular momentum operators, $\{ \hat{J}_{x}, \hat{J}_{y}, \hat{J}_{z}\}$ \cite{RodriguezLara2014,RodriguezLara2014b,Villanueva2015},
\begin{equation}
\mathbf{M} = \beta\, \mathbf{I} + \left[ \Delta \beta + f(\zeta) \right] \mathbf{J}_{z} + 2g\, \mathbf{J}_{x}, \label{eq:IniEq}
\end{equation}
where the propagation constant $\beta$ is common to all elements and the real parameter $\Delta \beta$ characterizes its variation due to the selected $SU(2)$ symmetry. 
The propagation dependent function $f(\zeta)$ summarizes the effect of controlled, small, periodic variations of individual propagation constants due to changes in refractive indices, resonator lengths, etc, depending on the specific realization. 
The parameter $g$ characterizes the coupling strength between nearest neighbor element-localized modes. 
%%%%%%%%%
Figure \ref{fig:1} displays an example of a waveguide array that generates the coupled mode matrix $\mathbf{M}$ in the three waveguide case. 
%%%%%%%%%
It is possible to use Wei-Norman factorization to calculate the propagation of modal amplitudes under the dynamics provided by this coupled mode matrix \cite{Wei1963,RogriguezLara2018}. 
However, we chose an alternate path of frame transformations for better understanding of the system behavior. 
In the following, we obviate the baseline constant term $\beta$ as it only induces an overall phase to the modal amplitudes vector.

The base system without periodic variations, $f(\zeta)=0$, constrains power transfer between elements due to the differences on individual effective propagation constants.
We can diagonalize the unmodulated coupled mode matrix, 
\begin{eqnarray}
\mathbf{M}_{0} = \Delta \beta\, \mathbf{J}_{z} + 2g\, \mathbf{J}_{x},
\end{eqnarray} 
using a rotation of the following form \cite{Villanueva2015}, 
\begin{eqnarray}
\mathbf{M}_{1}= e^{-i \theta\, \mathbf{J}_{y}} \cdot \mathbf{M}_{0} \cdot e^{i \theta\, \mathbf{J}_{y}}= \Omega\, \mathbf{J}_{z},
\end{eqnarray}
in terms of a rotation angle,
\begin{eqnarray}
\tan \theta = - \frac{2g}{\Delta \beta}.
\end{eqnarray}
The factor in the diagonal effective coupling matrix, defines an unmodulated device frequency,
\begin{equation}
\Omega = \sqrt{ \Delta \beta^2 + 4 g^{2} },
\end{equation} 
such that an array of identical elements, $\Delta \beta = 0$, produces complete power transfer between the $l$-th and the $(d-l+1)$-th elements at $\zeta = \pi / \Omega$ \cite{PerezLeija2013}. 
Differences on the individual refractive indices, $\Delta \beta \neq 0$, suppress power transfer \cite{Villanueva2015}.

Under $\zeta$-dependent modulation, this prescription no longer diagonalizes the coupled mode matrix,
\begin{eqnarray}
\mathbf{M}_{2} &=& e^{-i \theta\,\mathbf{J}_{y}} \cdot \mathbf{M} \cdot e^{i \theta\,\mathbf{J}_{y}} \nonumber \\
&=& \left[ \Omega + \frac{\Delta \beta\, f(\zeta)}{\Omega} \right] \mathbf{J}_{z} 
- \left[ \frac{2g\, f(\zeta)}{\Omega} \right] \mathbf{J}_{x}.
\end{eqnarray}
However, we can move to a $\zeta$-dependent frame defined by a rotation $e^{i \omega\, \mathbf{J}_{z} \, \zeta}$ to obtain an effective matrix, 
\begin{eqnarray}
\mathbf{M}_{3} 
&=& 
e^{-i \omega\, \mathbf{J}_{z}\, \zeta} \cdot 
\mathbf{M}_{2} \cdot 
e^{i \omega\, \mathbf{J}_{z}\, \zeta} 
- \omega\, \mathbf{J}_{z}, \nonumber \\
&=& 
\frac{\Delta \beta\, f(\zeta)}{\Omega}\, \mathbf{J}_{z} 
+ \left( \Omega - \omega \right) \mathbf{J}_{z} \nonumber \\
&&
- \frac{g\, f(\zeta)}{\Omega} \left( \mathbf{J}_{+}\, e^{-i \omega\, \zeta} + \mathbf{J}_{-}\, e^{i \omega\, \zeta}\right),
\end{eqnarray} 
where we used $\mathbf{J}_{x} = \left( \mathbf{J}_{+} + \mathbf{J}_{-} \right)/2$. 
This suggests the use of harmonic modulation,
\begin{equation}\label{Eq:Modulation}
f(\zeta) = \delta \sin( \omega\, \zeta ),
\end{equation}
to induce fast and slow frequencies in the system, 
\begin{eqnarray}
\mathbf{M}_{3} &=& 
\frac{\delta\, \Delta \beta}{\Omega} \sin(\omega\, \zeta)\, \mathbf{J}_{z} 
+ \left( \Omega - \omega \right) \mathbf{J}_{z} \nonumber \\
&&+ \frac{i \delta g}{2 \Omega} \!\left[ \left( 1 - e^{-i 2 \omega \zeta} \right) \mathbf{J}_{+} - \left(1 - e^{i 2 \omega \zeta} \right) \mathbf{J}_{-} \right]\!\!. \label{eq:BeforeRWA}
\end{eqnarray} 
Our system now includes five effective frequencies: $\delta\, \Delta \beta / \Omega$, $( \Omega-\omega) $, $\delta\, g / (2 \Omega)$, $\omega$ and $2 \omega$.
Control of the individual propagation constants by electro- or thermo-optic effects usually provides a small amplitude modulation parameter compared to the coupling strength and propagation constant difference, $\delta < g$ and $\delta < \Delta \beta$.
In consequence, it is smaller than the unmodulated device frequency $\delta \ll \Omega$. 
This leads to the following fact.
Unless we choose a modulation frequency close to the unmodulated device frequency, $\omega \sim \Omega$, the first and third terms on the right hand of Eq.~(\ref{eq:BeforeRWA}) may become irrelevant.
Furthermore, we can set them to be similar $\delta\, \Delta \beta / \Omega \sim \left(\Omega-\omega \right)\sim \delta\, g / (2 \Omega)$, leading to an ordered relation $\delta\, \Delta \beta / \Omega \sim \left(\Omega-\omega \right)\sim \delta\, g / (2 \Omega) \ll \omega < 2 \omega$.
This suggests using a rotating wave approximation (RWA) to neglect fast oscillating terms \cite{Allen1975},
\begin{equation}
\mathbf{M}_{3} \sim \left( \Omega - \omega \right) \mathbf{J}_{z} - \left( \frac{\delta\, g}{\Omega} \right) \mathbf{J}_{y},
\end{equation} 
and realize that we can diagonalize this effective coupled mode matrix by using a rotation along $\mathbf{J}_{x}$, 
\begin{eqnarray}
e^{- i \phi\, \mathbf{J}_{x}} \cdot \mathbf{M}_{3} \cdot e^{i \phi\, \mathbf{J}_{x}} = \Omega_{S}\, \mathbf{J}_{z},
\end{eqnarray} 
where the rotation angle $\phi$ fulfills the relation,
\begin{eqnarray}
\tan \phi = - \frac{\delta\, g}{\Omega \left( \Omega - \omega \right)}.
\end{eqnarray} 
This leads to an effective diagonal coupled mode matrix with a device frequency,
\begin{eqnarray}
\Omega_{S} = \sqrt{(\Omega-\omega)^{2}+\left(\frac{\delta g} { \Omega}\right)^{2}} \ll \Omega, \omega,
\end{eqnarray} 
slower than the frequencies of unmodulated device $\Omega$ and the thermal- or electro-optic modulation $\omega$. 
The modulated device provides complete power transfer between the $l$-th and the $(d-l+1)$-th waveguides for modulation frequency equal to the unmodulated device frequency, $\omega = \Omega$, at the approximate value $\zeta = \Omega\, \pi / (\delta\, g)$.

The rotating wave approximation, made by neglecting the fast oscillating terms with frequencies $\omega$ and $2\omega$, is equivalent to calculating the dynamics under a slowly varying envelope approximation (SVEA) of the modal field amplitude vector.
This can be seen in the solution for the modal amplitudes vector, 
\begin{eqnarray}
\vec{\mathcal{E}}(\zeta) 
&\sim&
e^{i \beta \zeta} \cdot 
e^{i \theta\, \mathbf{J}_{y}} \cdot 
e^{i \omega\, \mathbf{J}_{z}\, \zeta} \cdot 
e^{i \phi\, \mathbf{J}_{x}} \cdot 
e^{i \Omega_{S}\, \mathbf{J}_{z}\, \zeta} \cdot
e^{-i \phi\, \mathbf{J}_{x}} \cdot		
\nonumber \\ &&		
e^{-i \omega\, \mathbf{J}_{z}\, \zeta} \cdot 
e^{-i \theta\, \mathbf{J}_{y}} 
\cdot \vec{\mathcal{E}}(0),
\end{eqnarray}
where the initial state of the system is $\vec{\mathcal{E}}(0)$. 
It is straightforward to identify the $\zeta$-dependent terms responsible for the slowly varying envelope $e^{\pm i \Omega_{S}\, \mathbf{J}_{z}\, \zeta}$ and the fast modulation $e^{\pm i \omega\, \mathbf{J}_{z}\, \zeta}$ behaviors.

%%%%%%%%%%%%%%%%%%%%%%%%%%%%%%%%%%%%%%%%%%

\section{Telecom C-band example}

Let us consider a realization for telecom C-band, $\lambda = 1\,550~\mathrm{nm}$, using cylindrical fused silica waveguides. Since all waveguides must have different refractive indices, we use core refractive indices in the range $[1.445\,95, 1.447\,90]$ and a cladding index of $1.444\,00$ consistent with laser writing experiments \cite{Eaton2011}. 
We choose a core radius of $4.5~\mu\mathrm{m}$ to support a single LP$_{01}$ mode in each waveguide. 
For a realization with $d$ waveguides, the data above determines $\beta$ and $\Delta \beta$, fixing the localized single mode propagation constants for each waveguide. 
The corresponding individual core refractive indices can be numerically calculated from these. 
The next step is to find a reference value for the coupling parameter $g$. 
Doing so fixes the couplings between pairs of neighboring waveguides. 
The corresponding waveguide separations can be numerically extracted from this. 

\begin{figure*}[ht!]
	\centering
	\includegraphics[scale=0.9]{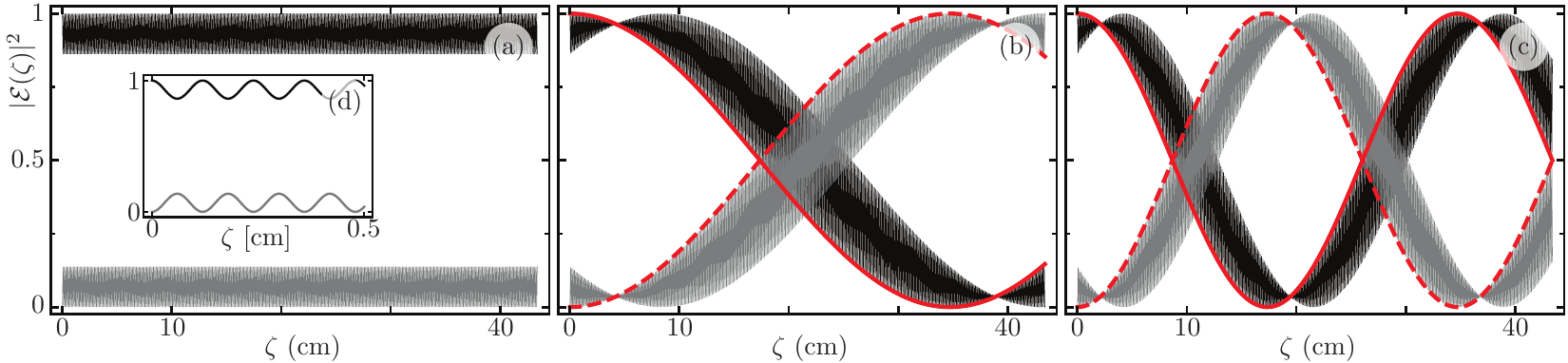}
	\caption{Squared field amplitude as a function of the propagation distance (a) with no refractive index modulation, (b) weak refractive index modulation, and (c) strong refractive index modulation. Inset (d) displays the squared field amplitude on a short propagation distance with limited power transfer oscillations. Black and gray curves display squared amplitudes within waveguides $l=1$ and $l=2$, respectively, provided by numerically solving the $\zeta$-dependent coupled matrix. Red solid and dashed curves are the analytic approximate slowly varying envelopes.}\label{fig:2}
\end{figure*}

Let us start with the simplest case, two waveguides described by the parameter set $\{ \beta, \Delta \beta, g \} = \{5.857\,32 \times 10^{6}, 4\,872.24, 974.447 \} ~\mathrm{rad}/\mathrm{m}$. 
The corresponding center-to-center core separation is $11.02~\mu\mathrm{m}$. 
We will assume thermo-optic effect as the drive behind periodic core refractive index modulation.
Silica first-order thermo-optic coefficient $1.090 \times 10^{-5}~\mathrm{K}^{-1}$ and temperature contrast of the order of 5$~\mathrm{K}$ between the cooler and the hotter parts of the waveguides yields an amplitude of the propagation constant modulation of the order $\delta = 97.444 \,7 ~\mathrm{rad}/\mathrm{m}$. 
Figure~\ref{fig:2}(a) displays the suppressed power-transfer between the two waveguides with no refractive index modulation. 
Figure~\ref{fig:2}(b) displays power transfer when the modulation period is on resonance, $\omega = \Omega$, and has an intermediate amplitude $\delta = 0.05 ~g = 48.722\,4~\mathrm{rad}/\mathrm{m}$. 
There, the distance for near complete power transfer is $\Omega\, \pi/(\delta\, g) = 34.72~\mathrm{cm}$.
Figure~\ref{fig:2}(c) displays power transfer on resonance with stronger modulation $\delta = 0.1 ~g = 97.444\,7 ~\mathrm{rad}/\mathrm{m}$ where the distance for near complete power transfer is $\Omega\, \pi/(\delta\, g) = 17.36~\mathrm{cm}$. As the modulation amplitude increases, the near complete transfer length decreases. 
The results in gray and black are numerical solutions for the full $\zeta$-dependent system without approximations, Eq.~(\ref{eq:IniEq}). 
The solid and dashed red lines are the evolution of just the approximate slowly varying envelope solution.

\begin{figure}[ht!]
	\centering
	\includegraphics{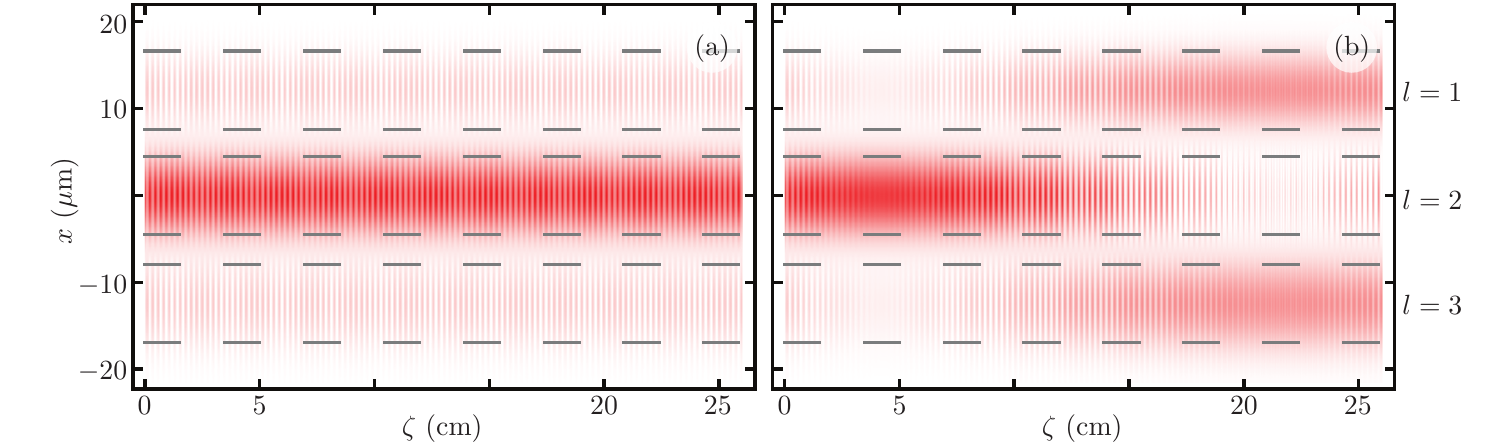}
	\caption{Squared field amplitude of the electric field in a coupled mode theory calculation with cylindrical waveguides. Here, light is impinged on the middle waveguide, $l=2$, (a) with and (b) without resonant, $\omega = \Omega$, refractive index modulation. The vertical axis, $x$, gives the separation between the waveguides and the horizontal axis, $\zeta$, is the propagation distance. More details in text.}\label{fig:3}
\end{figure}

Increasing the number of waveguides in the array allows for reconfigurable switching between more ports. 
As an example, we focus on the three waveguide case exemplified in Fig.~\ref{fig:1}, which yields a parameter set $\{ \beta, \Delta \beta, g \} = \{5.857\,32 \times 10^{6}, 2\,436.12, 487.224 \} ~\mathrm{rad}/\mathrm{m}$. 
The core refractive indices are now modulated to $\{ n_{1}, n_{2}, n_{3} \}= \{ 1.447\,90, 1.447\,02, 1.445\,95 \}$ and their separation distances $\{ d_{12}, d_{23} \} = \{12.44, 12.19\}~\mu\mathrm{m}$ are not equal to maintain a constant effective coupling with different refractive indices. 
The thermal contrast of 5$~\mathrm{K}$ yields a modulation amplitude $\delta = 0.1 ~g = 48.722\,4 ~\mathrm{rad}/\mathrm{m}$. 
The amplitude of the modulation $\delta$ is smaller than the one in the two-waveguide case but the maximum absolute change of individual core refractive indices are the same. 
Figure~\ref{fig:3} displays the squared field amplitude when light impinges on the central waveguide, $l=2$. 
Power transfer from the central core to the external ones occurs at $\Omega\, \pi / (2g\,\delta) = 17.36 ~\mathrm{cm}$. 
This plot contains coupled mode results, where the transversal information is obtained multiplying the coupled mode squared field amplitude results with the mode profiles of the waveguides, which are readily obtained with finite element analysis for the modes. 

\begin{figure}[ht!]
	\centering
	\includegraphics{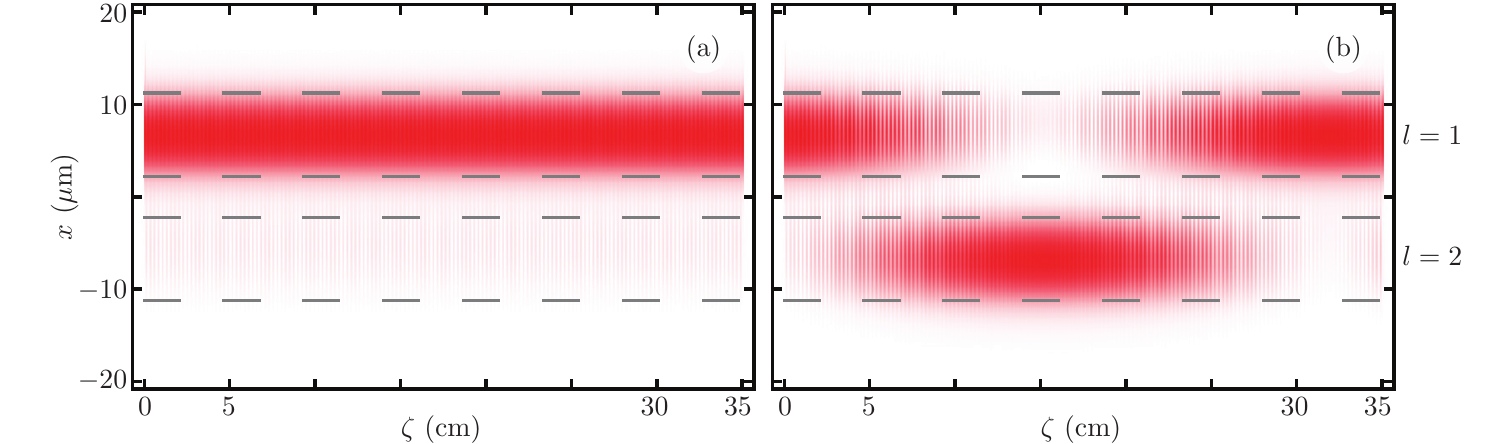}
	\caption{Squared field amplitude of the electric field in a 2D finite element simulation. Here, light is impinged on the upper waveguide, $l=1$, (a) with and (b) without resonant, $\omega = \Omega$, refractive index modulation. The vertical axis, $x$, gives the separation between the waveguides and the horizontal axis, $\zeta$, is the propagation distance. More details in text.}\label{fig:4}
\end{figure}

To insure the validity of our proposal, we go beyond the coupled mode theory approximation, and use a finite element simulation with COMSOL. 
In order to keep our simulation tractable in a standard workstation we restrict ourselves to two waveguides in a 2D simulation. 
This restriction is necessary due to the long propagation distances, of tens of centimeters. 
Our simulation describes planar silica waveguides in the telecom C-band with $9~\mu\mathrm{m}$ thickness and $13.45~\mu\mathrm{m}$ center-to-center separation, see Fig.~\ref{fig:4}. 
Their refractive indices are $1.447 90$ and $1.447 22$ for each of the upper and lower waveguides in Fig.~\ref{fig:4}(a) and (b), 
additionally the cladding and inter-waveguide refractive index is $1.444 00$. 
The unmodulated case, Fig.~\ref{fig:4}(a), displays limited power transfer between the waveguides. 
In contrast, the modulated case, Fig.~\ref{fig:4}(b), displays almost complete power transfer between the waveguides 
due to small refractive index changes along the propagation direction. 
The modulation has spatial frequency $\omega=2624.32~\mathrm{rad}/\mathrm{m}$ and amplitude of $3 \times 10^{-5}$, consistent with temperature differences $5~\mathrm{C}$ between the hottest and coldest parts of a waveguide.

%%%%%%%%%%%%%%%%%%%%%%%%%%%%%%%%%%%%%%%%%%
\section{Conclusions}

We demonstrated reconfigurable power transfer in nearest neighbor coupled mode systems by using periodic modulation of the localized modes effective propagation constant. 
In our proposal, an unmodulated system, where power transfer is heavily suppressed due to differences between the localized mode propagation constants, can show near complete power transfer if we introduce harmonic modulation with the correct frequency.
We provide an approximate effective power transfer frequency using the equivalent of a slowly varying envelope approximation for the modal field amplitudes. 
In our devices, the required modulation frequency is wavelength dependent through propagation constants and waveguide couplings. 
Therefore, unlike $\Delta \beta$ switches \cite{Kogelnik1975,Kogelnik1976,Schmidt1979,Korotky1986,Findakly1988,Tsukada1977,Molter1985}, to operate our device at different wavelengths it becomes necessary to adjust not only the modulation amplitude, but also the modulation frequency.

As a practical example, we work with fused silica waveguides and their first-order thermo-optic response in the telecom C-band. 
We present results using coupled mode theory and 2D finite element simulations. 
Here, the modulation amplitudes are small and the distances for power transfer become large, of the order of tens of centimeters. 
While proving the first principles validity of our proposal, this may not be feasible for photonic integrated devices. 
However, our results are valid for coupled mode systems, propagating in space or time, where the equivalent of localized propagation constants can be modulated in a controlled manner by electrical or thermal effects. 
For example, microring resonators or terahertz resonators, microwave cavities, radio frequency antennas, or RLC circuits.

%%%%%%%%%%%%%%%%%%%%%%%%%%%%%%%%%%%%%%%%%%

%%%%%%%%%%%%%%%%%%%%%%% References %%%%%%%%%%%%%%%%%%%%%%%%%
%%%%\bibliography{references}

\begin{thebibliography}{58}%
	\makeatletter
	\providecommand \@ifxundefined [1]{%
		\@ifx{#1\undefined}
	}%
	\providecommand \@ifnum [1]{%
		\ifnum #1\expandafter \@firstoftwo
		\else \expandafter \@secondoftwo
		\fi
	}%
	\providecommand \@ifx [1]{%
		\ifx #1\expandafter \@firstoftwo
		\else \expandafter \@secondoftwo
		\fi
	}%
	\providecommand \natexlab [1]{#1}%
	\providecommand \enquote  [1]{``#1''}%
	\providecommand \bibnamefont  [1]{#1}%
	\providecommand \bibfnamefont [1]{#1}%
	\providecommand \citenamefont [1]{#1}%
	\providecommand \href@noop [0]{\@secondoftwo}%
	\providecommand \href [0]{\begingroup \@sanitize@url \@href}%
	\providecommand \@href[1]{\@@startlink{#1}\@@href}%
	\providecommand \@@href[1]{\endgroup#1\@@endlink}%
	\providecommand \@sanitize@url [0]{\catcode `\\12\catcode `\$12\catcode
		`\&12\catcode `\#12\catcode `\^12\catcode `\_12\catcode `\%12\relax}%
	\providecommand \@@startlink[1]{}%
	\providecommand \@@endlink[0]{}%
	\providecommand \url  [0]{\begingroup\@sanitize@url \@url }%
	\providecommand \@url [1]{\endgroup\@href {#1}{\urlprefix }}%
	\providecommand \urlprefix  [0]{URL }%
	\providecommand \Eprint [0]{\href }%
	\providecommand \doibase [0]{http://dx.doi.org/}%
	\providecommand \selectlanguage [0]{\@gobble}%
	\providecommand \bibinfo  [0]{\@secondoftwo}%
	\providecommand \bibfield  [0]{\@secondoftwo}%
	\providecommand \translation [1]{[#1]}%
	\providecommand \BibitemOpen [0]{}%
	\providecommand \bibitemStop [0]{}%
	\providecommand \bibitemNoStop [0]{.\EOS\space}%
	\providecommand \EOS [0]{\spacefactor3000\relax}%
	\providecommand \BibitemShut  [1]{\csname bibitem#1\endcsname}%
	\let\auto@bib@innerbib\@empty
	%</preamble>
	\bibitem [{\citenamefont {Matthews}\ \emph {et~al.}(2009)\citenamefont
		{Matthews}, \citenamefont {Politi}, \citenamefont {Stefanov},\ and\
		\citenamefont {O'Brien}}]{Matthews2009}%
	\BibitemOpen
	\bibfield  {author} {\bibinfo {author} {\bibfnamefont {J.~C.~F.}\
			\bibnamefont {Matthews}}, \bibinfo {author} {\bibfnamefont {A.}~\bibnamefont
			{Politi}}, \bibinfo {author} {\bibfnamefont {A.}~\bibnamefont {Stefanov}}, \
		and\ \bibinfo {author} {\bibfnamefont {J.~L.}\ \bibnamefont {O'Brien}},\
	}\bibfield  {title} {\enquote {\bibinfo {title} {Manipulation of multiphoton
				entanglement in waveguide quantum circuits},}\ }\href {\doibase
		10.1038/nphoton.2009.93} {\bibfield  {journal} {\bibinfo  {journal} {Nat.
				Photonics}\ }\textbf {\bibinfo {volume} {3}},\ \bibinfo {pages} {346--350}
		(\bibinfo {year} {2009})},\ \Eprint {http://arxiv.org/abs/0911.1257}
	{arXiv:0911.1257 [quant-ph]} \BibitemShut {NoStop}%
	\bibitem [{\citenamefont {Carolan}\ \emph {et~al.}(2015)\citenamefont
		{Carolan}, \citenamefont {Harrold}, \citenamefont {Sparrow}, \citenamefont
		{Mart\'in-L\'opez}, \citenamefont {Russell}, \citenamefont {Silverstone},
		\citenamefont {Shadbolt}, \citenamefont {Matsuda}, \citenamefont {Oguma},
		\citenamefont {Itoh}, \citenamefont {Marshall}, \citenamefont {Thompson},
		\citenamefont {Matthews}, \citenamefont {Hashimoto}, \citenamefont
		{O'Brien},\ and\ \citenamefont {Laing}}]{Carolan2015}%
	\BibitemOpen
	\bibfield  {author} {\bibinfo {author} {\bibfnamefont {J.}~\bibnamefont
			{Carolan}}, \bibinfo {author} {\bibfnamefont {C.}~\bibnamefont {Harrold}},
		\bibinfo {author} {\bibfnamefont {C.}~\bibnamefont {Sparrow}}, \bibinfo
		{author} {\bibfnamefont {E.}~\bibnamefont {Mart\'in-L\'opez}}, \bibinfo
		{author} {\bibfnamefont {N.~J.}\ \bibnamefont {Russell}}, \bibinfo {author}
		{\bibfnamefont {J.~W.}\ \bibnamefont {Silverstone}}, \bibinfo {author}
		{\bibfnamefont {P.~J.}\ \bibnamefont {Shadbolt}}, \bibinfo {author}
		{\bibfnamefont {N.}~\bibnamefont {Matsuda}}, \bibinfo {author} {\bibfnamefont
			{M.}~\bibnamefont {Oguma}}, \bibinfo {author} {\bibfnamefont
			{M.}~\bibnamefont {Itoh}}, \bibinfo {author} {\bibfnamefont {G.~D.}\
			\bibnamefont {Marshall}}, \bibinfo {author} {\bibfnamefont {M.~G.}\
			\bibnamefont {Thompson}}, \bibinfo {author} {\bibfnamefont {J.~C.~F.}\
			\bibnamefont {Matthews}}, \bibinfo {author} {\bibfnamefont {T.}~\bibnamefont
			{Hashimoto}}, \bibinfo {author} {\bibfnamefont {J.~L.}\ \bibnamefont
			{O'Brien}}, \ and\ \bibinfo {author} {\bibfnamefont {A.}~\bibnamefont
			{Laing}},\ }\bibfield  {title} {\enquote {\bibinfo {title} {Universal linear
				optics},}\ }\href {\doibase 10.1126/science.aab3642} {\bibfield  {journal}
		{\bibinfo  {journal} {Science}\ }\textbf {\bibinfo {volume} {349}},\ \bibinfo
		{pages} {711--716} (\bibinfo {year} {2015})},\ \Eprint
	{http://arxiv.org/abs/1505.01182} {arXiv:1505.01182 [quant-ph]} \BibitemShut
	{NoStop}%
	\bibitem [{\citenamefont {Wang}\ \emph {et~al.}(2018)\citenamefont {Wang},
		\citenamefont {Paesani}, \citenamefont {Ding}, \citenamefont {Santagati},
		\citenamefont {Skrzypczyk}, \citenamefont {Salavrakos}, \citenamefont {Tura},
		\citenamefont {Augusiak}, \citenamefont {Man{\v c}inska}, \citenamefont
		{Bacco}, \citenamefont {Bonneau}, \citenamefont {Silverstone}, \citenamefont
		{Gong}, \citenamefont {Ac\'in}, \citenamefont {Rottwitt}, \citenamefont
		{Oxenl{\o}we}, \citenamefont {O'Brien}, \citenamefont {Laing},\ and\
		\citenamefont {Thompson}}]{Wang2019}%
	\BibitemOpen
	\bibfield  {author} {\bibinfo {author} {\bibfnamefont {J.}~\bibnamefont
			{Wang}}, \bibinfo {author} {\bibfnamefont {S.}~\bibnamefont {Paesani}},
		\bibinfo {author} {\bibfnamefont {Y.}~\bibnamefont {Ding}}, \bibinfo {author}
		{\bibfnamefont {R.}~\bibnamefont {Santagati}}, \bibinfo {author}
		{\bibfnamefont {P.}~\bibnamefont {Skrzypczyk}}, \bibinfo {author}
		{\bibfnamefont {A.}~\bibnamefont {Salavrakos}}, \bibinfo {author}
		{\bibfnamefont {J.}~\bibnamefont {Tura}}, \bibinfo {author} {\bibfnamefont
			{R.}~\bibnamefont {Augusiak}}, \bibinfo {author} {\bibfnamefont
			{L.}~\bibnamefont {Man{\v c}inska}}, \bibinfo {author} {\bibfnamefont
			{D.}~\bibnamefont {Bacco}}, \bibinfo {author} {\bibfnamefont
			{D.}~\bibnamefont {Bonneau}}, \bibinfo {author} {\bibfnamefont {J.~W.}\
			\bibnamefont {Silverstone}}, \bibinfo {author} {\bibfnamefont
			{O.}~\bibnamefont {Gong}}, \bibinfo {author} {\bibfnamefont {A.}~\bibnamefont
			{Ac\'in}}, \bibinfo {author} {\bibfnamefont {K.}~\bibnamefont {Rottwitt}},
		\bibinfo {author} {\bibfnamefont {L.~K.}\ \bibnamefont {Oxenl{\o}we}},
		\bibinfo {author} {\bibfnamefont {J.~L.}\ \bibnamefont {O'Brien}}, \bibinfo
		{author} {\bibfnamefont {A.}~\bibnamefont {Laing}}, \ and\ \bibinfo {author}
		{\bibfnamefont {M.~G.}\ \bibnamefont {Thompson}},\ }\bibfield  {title}
	{\enquote {\bibinfo {title} {Multidimensional quantum entanglement with
				large-scale integrated optics},}\ }\href {\doibase 10.1126/science.aar7053}
	{\bibfield  {journal} {\bibinfo  {journal} {Science}\ }\textbf {\bibinfo
			{volume} {360}},\ \bibinfo {pages} {285--291} (\bibinfo {year} {2018})},\
	\Eprint {http://arxiv.org/abs/1803.04449} {arXiv:1803.04449 [quant-ph]}
	\BibitemShut {NoStop}%
	\bibitem [{\citenamefont {Giles}\ \emph {et~al.}(1999)\citenamefont {Giles},
		\citenamefont {Aksyuk}, \citenamefont {Barber}, \citenamefont {Ruel},
		\citenamefont {Stulz},\ and\ \citenamefont {Bishop}}]{Giles1999}%
	\BibitemOpen
	\bibfield  {author} {\bibinfo {author} {\bibfnamefont {C.~R.}\ \bibnamefont
			{Giles}}, \bibinfo {author} {\bibfnamefont {V.}~\bibnamefont {Aksyuk}},
		\bibinfo {author} {\bibfnamefont {B.}~\bibnamefont {Barber}}, \bibinfo
		{author} {\bibfnamefont {R.}~\bibnamefont {Ruel}}, \bibinfo {author}
		{\bibfnamefont {L.}~\bibnamefont {Stulz}}, \ and\ \bibinfo {author}
		{\bibfnamefont {D.}~\bibnamefont {Bishop}},\ }\bibfield  {title} {\enquote
		{\bibinfo {title} {A silicon {MEMS} optical switch attenuator and its use in
				lightwave subsystems},}\ }\href {\doibase 10.1109/2944.748100} {\bibfield
		{journal} {\bibinfo  {journal} {IEEE J. Sel. Top. Quant.}\ }\textbf {\bibinfo
			{volume} {5}},\ \bibinfo {pages} {18--25} (\bibinfo {year}
		{1999})}\BibitemShut {NoStop}%
	\bibitem [{\citenamefont {Borovic}\ \emph {et~al.}(2004)\citenamefont
		{Borovic}, \citenamefont {Hong}, \citenamefont {Liu}, \citenamefont {Xie},\
		and\ \citenamefont {Lewis}}]{Borovic2004}%
	\BibitemOpen
	\bibfield  {author} {\bibinfo {author} {\bibfnamefont {B.}~\bibnamefont
			{Borovic}}, \bibinfo {author} {\bibfnamefont {C.}~\bibnamefont {Hong}},
		\bibinfo {author} {\bibfnamefont {A.~Q.}\ \bibnamefont {Liu}}, \bibinfo
		{author} {\bibfnamefont {L.}~\bibnamefont {Xie}}, \ and\ \bibinfo {author}
		{\bibfnamefont {F.~L.}\ \bibnamefont {Lewis}},\ }\bibfield  {title} {\enquote
		{\bibinfo {title} {Control of a {MEMS} optical switch},}\ }in\ \href
	{\doibase 10.1109/CDC.2004.1428930} {\emph {\bibinfo {booktitle} {2004 43rd
				{IEEE} Conference on Decision and Control ({CDC})}}},\ Vol.~\bibinfo {volume}
	{3}\ (\bibinfo {year} {2004})\ pp.\ \bibinfo {pages} {3039--3044}\BibitemShut
	{NoStop}%
	\bibitem [{\citenamefont {Yano}\ \emph {et~al.}(2005)\citenamefont {Yano},
		\citenamefont {Yamagishi},\ and\ \citenamefont {Tsuda}}]{Yano2005}%
	\BibitemOpen
	\bibfield  {author} {\bibinfo {author} {\bibfnamefont {M.}~\bibnamefont
			{Yano}}, \bibinfo {author} {\bibfnamefont {F.}~\bibnamefont {Yamagishi}}, \
		and\ \bibinfo {author} {\bibfnamefont {T.}~\bibnamefont {Tsuda}},\ }\bibfield
	{title} {\enquote {\bibinfo {title} {Optical {MEMS} for photonic
				switching-compact and stable optical crossconnect switches for simple, fast,
				and flexible wavelength applications in recent photonic networks},}\ }\href
	{\doibase 10.1109/JSTQE.2005.846523} {\bibfield  {journal} {\bibinfo
			{journal} {IEEE J. Sel. Top. Quant.}\ }\textbf {\bibinfo {volume} {11}},\
		\bibinfo {pages} {383--394} (\bibinfo {year} {2005})}\BibitemShut {NoStop}%
	\bibitem [{\citenamefont {Tsybeskov}\ \emph {et~al.}(2009)\citenamefont
		{Tsybeskov}, \citenamefont {Lockwood},\ and\ \citenamefont
		{Ichikawa}}]{Tsybeskov2009}%
	\BibitemOpen
	\bibfield  {author} {\bibinfo {author} {\bibfnamefont {L.}~\bibnamefont
			{Tsybeskov}}, \bibinfo {author} {\bibfnamefont {D.~J.}\ \bibnamefont
			{Lockwood}}, \ and\ \bibinfo {author} {\bibfnamefont {M.}~\bibnamefont
			{Ichikawa}},\ }\bibfield  {title} {\enquote {\bibinfo {title} {Silicon
				photonics: {CMOS} going optical [scanning the issue]},}\ }\href {\doibase
		10.1109/JPROC.2009.2021052} {\bibfield  {journal} {\bibinfo  {journal} {Proc.
				IEEE}\ }\textbf {\bibinfo {volume} {97}},\ \bibinfo {pages} {1161--1165}
		(\bibinfo {year} {2009})}\BibitemShut {NoStop}%
	\bibitem [{\citenamefont {Rylyakov}\ \emph {et~al.}(2012)\citenamefont
		{Rylyakov}, \citenamefont {Schow}, \citenamefont {Lee}, \citenamefont
		{Green}, \citenamefont {Assefa}, \citenamefont {Doany}, \citenamefont {Yang},
		\citenamefont {{Van Campenhout}}, \citenamefont {Jahnes}, \citenamefont
		{Kash},\ and\ \citenamefont {Vlasov}}]{Rylyakov2012}%
	\BibitemOpen
	\bibfield  {author} {\bibinfo {author} {\bibfnamefont {A.~V.}\ \bibnamefont
			{Rylyakov}}, \bibinfo {author} {\bibfnamefont {C.~L.}\ \bibnamefont {Schow}},
		\bibinfo {author} {\bibfnamefont {B.~G.}\ \bibnamefont {Lee}}, \bibinfo
		{author} {\bibfnamefont {W.~M.~J.}\ \bibnamefont {Green}}, \bibinfo {author}
		{\bibfnamefont {S.}~\bibnamefont {Assefa}}, \bibinfo {author} {\bibfnamefont
			{F.~E.}\ \bibnamefont {Doany}}, \bibinfo {author} {\bibfnamefont
			{M.}~\bibnamefont {Yang}}, \bibinfo {author} {\bibfnamefont {J.}~\bibnamefont
			{{Van Campenhout}}}, \bibinfo {author} {\bibfnamefont {C.~V.}\ \bibnamefont
			{Jahnes}}, \bibinfo {author} {\bibfnamefont {J.~A.}\ \bibnamefont {Kash}}, \
		and\ \bibinfo {author} {\bibfnamefont {Y.~A.}\ \bibnamefont {Vlasov}},\
	}\bibfield  {title} {\enquote {\bibinfo {title} {Silicon photonic switches
				hybrid-integrated with {CMOS} drivers},}\ }\href {\doibase
		10.1109/JSSC.2011.2170638} {\bibfield  {journal} {\bibinfo  {journal} {IEEE
				J. Solid-St. Circ.}\ }\textbf {\bibinfo {volume} {47}},\ \bibinfo {pages}
		{345--354} (\bibinfo {year} {2012})}\BibitemShut {NoStop}%
	\bibitem [{\citenamefont {Longhi}(2009)}]{Longhi2009}%
	\BibitemOpen
	\bibfield  {author} {\bibinfo {author} {\bibfnamefont {S.}~\bibnamefont
			{Longhi}},\ }\bibfield  {title} {\enquote {\bibinfo {title} {Quantum-optical
				analogies using photonic structures},}\ }\href {\doibase
		10.1002/lpor.200810055} {\bibfield  {journal} {\bibinfo  {journal} {Laser
				Photonics Rev.}\ }\textbf {\bibinfo {volume} {3}},\ \bibinfo {pages}
		{243--261} (\bibinfo {year} {2009})}\BibitemShut {NoStop}%
	\bibitem [{\citenamefont {Rodr\'iguez-Lara}\ and\ \citenamefont
		{Guerrero}(2015)}]{RodriguezLara2015b}%
	\BibitemOpen
	\bibfield  {author} {\bibinfo {author} {\bibfnamefont {B.~M.}\ \bibnamefont
			{Rodr\'iguez-Lara}}\ and\ \bibinfo {author} {\bibfnamefont {J.}~\bibnamefont
			{Guerrero}},\ }\bibfield  {title} {\enquote {\bibinfo {title} {Optical finite
				representation of the {L}orentz group},}\ }\href {\doibase
		10.1364/OL.40.005682} {\bibfield  {journal} {\bibinfo  {journal} {Opt.
				Lett.}\ }\textbf {\bibinfo {volume} {40}},\ \bibinfo {pages} {5682--5685}
		(\bibinfo {year} {2015})},\ \Eprint {http://arxiv.org/abs/1508.05419}
	{arXiv:1508.05419 [physics.optics]} \BibitemShut {NoStop}%
	\bibitem [{\citenamefont {{Villanueva Vergara}}\ and\ \citenamefont
		{Rodr\'iguez-Lara}(2015)}]{Villanueva2015}%
	\BibitemOpen
	\bibfield  {author} {\bibinfo {author} {\bibfnamefont {L.}~\bibnamefont
			{{Villanueva Vergara}}}\ and\ \bibinfo {author} {\bibfnamefont {B.~M.}\
			\bibnamefont {Rodr\'iguez-Lara}},\ }\bibfield  {title} {\enquote {\bibinfo
			{title} {{G}ilmore-{P}erelomov symmetry based approach to photonic
				lattices},}\ }\href {\doibase 10.1364/OE.23.022836} {\bibfield  {journal}
		{\bibinfo  {journal} {Opt. Express}\ }\textbf {\bibinfo {volume} {23}},\
		\bibinfo {pages} {22836--22846} (\bibinfo {year} {2015})},\ \Eprint
	{http://arxiv.org/abs/1506.02062} {arXiv:1506.02062 [physics.optics]}
	\BibitemShut {NoStop}%
	\bibitem [{\citenamefont {Rodríguez-Lara}\ \emph {et~al.}(2018)\citenamefont
		{Rodríguez-Lara}, \citenamefont {El-Ganainy},\ and\ \citenamefont
		{Guerrero}}]{RogriguezLara2018}%
	\BibitemOpen
	\bibfield  {author} {\bibinfo {author} {\bibfnamefont {B.~M.}\ \bibnamefont
			{Rodríguez-Lara}}, \bibinfo {author} {\bibfnamefont {R.}~\bibnamefont
			{El-Ganainy}}, \ and\ \bibinfo {author} {\bibfnamefont {J.}~\bibnamefont
			{Guerrero}},\ }\bibfield  {title} {\enquote {\bibinfo {title} {Symmetry in
				optics and photonics: a group theory approach},}\ }\href {\doibase
		10.1016/j.scib.2017.12.020} {\bibfield  {journal} {\bibinfo  {journal} {Sci.
				Bull.}\ }\textbf {\bibinfo {volume} {63}},\ \bibinfo {pages} {244--251}
		(\bibinfo {year} {2018})},\ \Eprint {http://arxiv.org/abs/1803.00121}
	{arXiv:1803.00121 [physics.optics]} \BibitemShut {NoStop}%
	\bibitem [{\citenamefont {Longhi}(2006)}]{Longhi2006}%
	\BibitemOpen
	\bibfield  {author} {\bibinfo {author} {\bibfnamefont {S.}~\bibnamefont
			{Longhi}},\ }\bibfield  {title} {\enquote {\bibinfo {title} {Adiabatic
				passage of light in coupled optical waveguides},}\ }\href {\doibase
		10.1103/PhysRevE.73.026607} {\bibfield  {journal} {\bibinfo  {journal} {Phys.
				Rev. E}\ }\textbf {\bibinfo {volume} {73}},\ \bibinfo {pages} {026607}
		(\bibinfo {year} {2006})}\BibitemShut {NoStop}%
	\bibitem [{\citenamefont {Longhi}\ \emph {et~al.}(2007)\citenamefont {Longhi},
		\citenamefont {{Della Valle}}, \citenamefont {Ornigotti},\ and\ \citenamefont
		{Laporta}}]{Longhi2007}%
	\BibitemOpen
	\bibfield  {author} {\bibinfo {author} {\bibfnamefont {S.}~\bibnamefont
			{Longhi}}, \bibinfo {author} {\bibfnamefont {G.}~\bibnamefont {{Della
					Valle}}}, \bibinfo {author} {\bibfnamefont {M.}~\bibnamefont {Ornigotti}}, \
		and\ \bibinfo {author} {\bibfnamefont {P.}~\bibnamefont {Laporta}},\
	}\bibfield  {title} {\enquote {\bibinfo {title} {Coherent tunneling by
				adiabatic passage in an optical waveguide system},}\ }\href {\doibase
		10.1103/PhysRevB.76.201101} {\bibfield  {journal} {\bibinfo  {journal} {Phys.
				Rev. B}\ }\textbf {\bibinfo {volume} {76}},\ \bibinfo {pages} {201101(R)}
		(\bibinfo {year} {2007})},\ \Eprint {http://arxiv.org/abs/0709.3050}
	{arXiv:0709.3050 [cond-mat.other]} \BibitemShut {NoStop}%
	\bibitem [{\citenamefont {{Della Valle}}\ \emph {et~al.}(2008)\citenamefont
		{{Della Valle}}, \citenamefont {Ornigotti}, \citenamefont {{Toney
				Fernandez}}, \citenamefont {Laporta}, \citenamefont {Longhi}, \citenamefont
		{Coppa},\ and\ \citenamefont {Foglietti}}]{Della2008}%
	\BibitemOpen
	\bibfield  {author} {\bibinfo {author} {\bibfnamefont {G.}~\bibnamefont
			{{Della Valle}}}, \bibinfo {author} {\bibfnamefont {M.}~\bibnamefont
			{Ornigotti}}, \bibinfo {author} {\bibfnamefont {T.}~\bibnamefont {{Toney
					Fernandez}}}, \bibinfo {author} {\bibfnamefont {P.}~\bibnamefont {Laporta}},
		\bibinfo {author} {\bibfnamefont {S.}~\bibnamefont {Longhi}}, \bibinfo
		{author} {\bibfnamefont {A.}~\bibnamefont {Coppa}}, \ and\ \bibinfo {author}
		{\bibfnamefont {V.}~\bibnamefont {Foglietti}},\ }\bibfield  {title} {\enquote
		{\bibinfo {title} {Adiabatic light transfer via dressed states in optical
				waveguide arrays},}\ }\href {\doibase 10.1063/1.2828985} {\bibfield
		{journal} {\bibinfo  {journal} {Appl. Phys. Lett.}\ }\textbf {\bibinfo
			{volume} {92}},\ \bibinfo {pages} {011106} (\bibinfo {year}
		{2008})}\BibitemShut {NoStop}%
	\bibitem [{\citenamefont {Perez-Leija}\ \emph {et~al.}(2013)\citenamefont
		{Perez-Leija}, \citenamefont {Keil}, \citenamefont {Kay}, \citenamefont
		{Moya-Cessa}, \citenamefont {Nolte}, \citenamefont {Kwek}, \citenamefont
		{Rodr\'iguez-Lara}, \citenamefont {Szameit},\ and\ \citenamefont
		{Christodoulides}}]{PerezLeija2013}%
	\BibitemOpen
	\bibfield  {author} {\bibinfo {author} {\bibfnamefont {A.}~\bibnamefont
			{Perez-Leija}}, \bibinfo {author} {\bibfnamefont {R.}~\bibnamefont {Keil}},
		\bibinfo {author} {\bibfnamefont {A.}~\bibnamefont {Kay}}, \bibinfo {author}
		{\bibfnamefont {H.}~\bibnamefont {Moya-Cessa}}, \bibinfo {author}
		{\bibfnamefont {S.}~\bibnamefont {Nolte}}, \bibinfo {author} {\bibfnamefont
			{L.-C.}\ \bibnamefont {Kwek}}, \bibinfo {author} {\bibfnamefont {B.~M.}\
			\bibnamefont {Rodr\'iguez-Lara}}, \bibinfo {author} {\bibfnamefont
			{A.}~\bibnamefont {Szameit}}, \ and\ \bibinfo {author} {\bibfnamefont
			{D.~N.}\ \bibnamefont {Christodoulides}},\ }\bibfield  {title} {\enquote
		{\bibinfo {title} {Coherent quantum transport in photonic lattices},}\ }\href
	{\doibase 10.1103/PhysRevA.87.012309} {\bibfield  {journal} {\bibinfo
			{journal} {Phys. Rev. A}\ }\textbf {\bibinfo {volume} {87}},\ \bibinfo
		{pages} {012309} (\bibinfo {year} {2013})},\ \Eprint
	{http://arxiv.org/abs/1207.6080} {arXiv:1207.6080 [quant-ph]} \BibitemShut
	{NoStop}%
	\bibitem [{\citenamefont {John}(1984)}]{John1984}%
	\BibitemOpen
	\bibfield  {author} {\bibinfo {author} {\bibfnamefont {S.}~\bibnamefont
			{John}},\ }\bibfield  {title} {\enquote {\bibinfo {title} {Electromagnetic
				absorption in a disordered medium near a photon mobility edge},}\ }\href
	{\doibase 10.1103/PhysRevLett.53.2169} {\bibfield  {journal} {\bibinfo
			{journal} {Phys. Rev. Lett.}\ }\textbf {\bibinfo {volume} {53}},\ \bibinfo
		{pages} {2169--2172} (\bibinfo {year} {1984})}\BibitemShut {NoStop}%
	\bibitem [{\citenamefont {{De Raedt}}\ \emph {et~al.}(1989)\citenamefont {{De
				Raedt}}, \citenamefont {Lagendijk},\ and\ \citenamefont {{de
				Vries}}}]{DeRaedt1989}%
	\BibitemOpen
	\bibfield  {author} {\bibinfo {author} {\bibfnamefont {H.}~\bibnamefont {{De
					Raedt}}}, \bibinfo {author} {\bibfnamefont {A.}~\bibnamefont {Lagendijk}}, \
		and\ \bibinfo {author} {\bibfnamefont {P.}~\bibnamefont {{de Vries}}},\
	}\bibfield  {title} {\enquote {\bibinfo {title} {Transverse localization of
				light},}\ }\href {\doibase 10.1103/PhysRevLett.62.47} {\bibfield  {journal}
		{\bibinfo  {journal} {Phys. Rev. Lett.}\ }\textbf {\bibinfo {volume} {62}},\
		\bibinfo {pages} {47--50} (\bibinfo {year} {1989})}\BibitemShut {NoStop}%
	\bibitem [{\citenamefont {Schwartz}\ \emph {et~al.}(2007)\citenamefont
		{Schwartz}, \citenamefont {Bartal}, \citenamefont {Fishman},\ and\
		\citenamefont {Segev}}]{Schwartz2007}%
	\BibitemOpen
	\bibfield  {author} {\bibinfo {author} {\bibfnamefont {T.}~\bibnamefont
			{Schwartz}}, \bibinfo {author} {\bibfnamefont {G.}~\bibnamefont {Bartal}},
		\bibinfo {author} {\bibfnamefont {S.}~\bibnamefont {Fishman}}, \ and\
		\bibinfo {author} {\bibfnamefont {M.}~\bibnamefont {Segev}},\ }\bibfield
	{title} {\enquote {\bibinfo {title} {Transport and {A}nderson localization in
				disordered two-dimensional photonic lattices},}\ }\href {\doibase
		10.1038/nature05623} {\bibfield  {journal} {\bibinfo  {journal} {Nature}\
		}\textbf {\bibinfo {volume} {446}},\ \bibinfo {pages} {52--55} (\bibinfo
		{year} {2007})}\BibitemShut {NoStop}%
	\bibitem [{\citenamefont {Segev}\ \emph {et~al.}(2013)\citenamefont {Segev},
		\citenamefont {Silberberg},\ and\ \citenamefont
		{Christodoulides}}]{Segev2013}%
	\BibitemOpen
	\bibfield  {author} {\bibinfo {author} {\bibfnamefont {M.}~\bibnamefont
			{Segev}}, \bibinfo {author} {\bibfnamefont {Y.}~\bibnamefont {Silberberg}}, \
		and\ \bibinfo {author} {\bibfnamefont {D.~N.}\ \bibnamefont
			{Christodoulides}},\ }\bibfield  {title} {\enquote {\bibinfo {title}
			{{A}nderson localization of light},}\ }\href {\doibase
		10.1038/nphoton.2013.30} {\bibfield  {journal} {\bibinfo  {journal} {Nat.
				Photonics}\ }\textbf {\bibinfo {volume} {7}},\ \bibinfo {pages} {197--204}
		(\bibinfo {year} {2013})}\BibitemShut {NoStop}%
	\bibitem [{\citenamefont {{Jaramillo \'Avila}}\ \emph
		{et~al.}(2019)\citenamefont {{Jaramillo \'Avila}}, \citenamefont {Torres},
		\citenamefont {de~J.~Le\'on-Montiel},\ and\ \citenamefont
		{Rodr\'iguez-Lara}}]{Jaramillo2019}%
	\BibitemOpen
	\bibfield  {author} {\bibinfo {author} {\bibfnamefont {B.}~\bibnamefont
			{{Jaramillo \'Avila}}}, \bibinfo {author} {\bibfnamefont {J.~M.}\
			\bibnamefont {Torres}}, \bibinfo {author} {\bibfnamefont {R.}~\bibnamefont
			{de~J.~Le\'on-Montiel}}, \ and\ \bibinfo {author} {\bibfnamefont {B.~M.}\
			\bibnamefont {Rodr\'iguez-Lara}},\ }\bibfield  {title} {\enquote {\bibinfo
			{title} {Optimal crosstalk suppression in multicore fibers},}\ }\href
	{\doibase 10.1038/s41598-019-51854-x} {\bibfield  {journal} {\bibinfo
			{journal} {Sci. Rep.}\ }\textbf {\bibinfo {volume} {9}},\ \bibinfo {pages}
		{15737} (\bibinfo {year} {2019})},\ \Eprint {http://arxiv.org/abs/1905.09416}
	{arXiv:1905.09416 [physics.optics]} \BibitemShut {NoStop}%
	\bibitem [{\citenamefont {Ruschhaupt}\ \emph {et~al.}(2005)\citenamefont
		{Ruschhaupt}, \citenamefont {Delgado},\ and\ \citenamefont
		{Muga}}]{Ruschhaupt2005}%
	\BibitemOpen
	\bibfield  {author} {\bibinfo {author} {\bibfnamefont {A.}~\bibnamefont
			{Ruschhaupt}}, \bibinfo {author} {\bibfnamefont {F.}~\bibnamefont {Delgado}},
		\ and\ \bibinfo {author} {\bibfnamefont {J.~G.}\ \bibnamefont {Muga}},\
	}\bibfield  {title} {\enquote {\bibinfo {title} {Physical realization of
				{PT}-symmetric potential scattering in a planar slab waveguide},}\ }\href
	{\doibase 10.1088/0305-4470/38/9/L03} {\bibfield  {journal} {\bibinfo
			{journal} {J. Phys. A: Math. Gen}\ }\textbf {\bibinfo {volume} {38}},\
		\bibinfo {pages} {L171--L176} (\bibinfo {year} {2005})}\BibitemShut {NoStop}%
	\bibitem [{\citenamefont {El-Ganainy}\ \emph {et~al.}(2007)\citenamefont
		{El-Ganainy}, \citenamefont {Makris}, \citenamefont {Christodoulides},\ and\
		\citenamefont {Musslimani}}]{ElGanainy2007}%
	\BibitemOpen
	\bibfield  {author} {\bibinfo {author} {\bibfnamefont {R.}~\bibnamefont
			{El-Ganainy}}, \bibinfo {author} {\bibfnamefont {K.~G.}\ \bibnamefont
			{Makris}}, \bibinfo {author} {\bibfnamefont {D.~N.}\ \bibnamefont
			{Christodoulides}}, \ and\ \bibinfo {author} {\bibfnamefont {Z.~H.}\
			\bibnamefont {Musslimani}},\ }\bibfield  {title} {\enquote {\bibinfo {title}
			{Theory of coupled optical {PT}-symmetric structures},}\ }\href {\doibase
		10.1364/OL.32.002632} {\bibfield  {journal} {\bibinfo  {journal} {Opt.
				Lett.}\ }\textbf {\bibinfo {volume} {32}},\ \bibinfo {pages} {2632--2634}
		(\bibinfo {year} {2007})}\BibitemShut {NoStop}%
	\bibitem [{\citenamefont {{Huerta Morales}}\ \emph {et~al.}(2016)\citenamefont
		{{Huerta Morales}}, \citenamefont {Guerrero}, \citenamefont
		{L\'opez-Aguayo},\ and\ \citenamefont
		{Rodr\'iguez-Lara}}]{HuertaMorales2016}%
	\BibitemOpen
	\bibfield  {author} {\bibinfo {author} {\bibfnamefont {J.~D.}\ \bibnamefont
			{{Huerta Morales}}}, \bibinfo {author} {\bibfnamefont {J.}~\bibnamefont
			{Guerrero}}, \bibinfo {author} {\bibfnamefont {S.}~\bibnamefont
			{L\'opez-Aguayo}}, \ and\ \bibinfo {author} {\bibfnamefont {B.~M.}\
			\bibnamefont {Rodr\'iguez-Lara}},\ }\bibfield  {title} {\enquote {\bibinfo
			{title} {Revisiting the optical {PT}-symmetric dimer},}\ }\href {\doibase
		10.3390/sym8090083} {\bibfield  {journal} {\bibinfo  {journal} {Symmetry}\
		}\textbf {\bibinfo {volume} {8}},\ \bibinfo {pages} {83} (\bibinfo {year}
		{2016})},\ \Eprint {http://arxiv.org/abs/1607.02782} {arXiv:1607.02782
		[physics.optics]} \BibitemShut {NoStop}%
	\bibitem [{\citenamefont {{Nodal Stevens}}\ \emph {et~al.}(2018)\citenamefont
		{{Nodal Stevens}}, \citenamefont {{Jaramillo \'Avila}},\ and\ \citenamefont
		{Rodr\'iguez-Lara}}]{Nodal2018}%
	\BibitemOpen
	\bibfield  {author} {\bibinfo {author} {\bibfnamefont {D.~J.}\ \bibnamefont
			{{Nodal Stevens}}}, \bibinfo {author} {\bibfnamefont {B.}~\bibnamefont
			{{Jaramillo \'Avila}}}, \ and\ \bibinfo {author} {\bibfnamefont {B.~M.}\
			\bibnamefont {Rodr\'iguez-Lara}},\ }\bibfield  {title} {\enquote {\bibinfo
			{title} {Necklaces of {PT}-symmetric dimers},}\ }\href {\doibase
		10.1364/PRJ.6.000A31} {\bibfield  {journal} {\bibinfo  {journal} {Photon.
				Res.}\ }\textbf {\bibinfo {volume} {6}},\ \bibinfo {pages} {A31--A37}
		(\bibinfo {year} {2018})},\ \Eprint {http://arxiv.org/abs/1709.00498}
	{arXiv:1709.00498 [physics.optics]} \BibitemShut {NoStop}%
	\bibitem [{\citenamefont {Rodr\'iguez-Lara}\ \emph
		{et~al.}(2014{\natexlab{a}})\citenamefont {Rodr\'iguez-Lara}, \citenamefont
		{Aleahmad}, \citenamefont {Moya-Cessa},\ and\ \citenamefont
		{Christodoulides}}]{RodriguezLara2014}%
	\BibitemOpen
	\bibfield  {author} {\bibinfo {author} {\bibfnamefont {B.~M.}\ \bibnamefont
			{Rodr\'iguez-Lara}}, \bibinfo {author} {\bibfnamefont {P.}~\bibnamefont
			{Aleahmad}}, \bibinfo {author} {\bibfnamefont {H.~M.}\ \bibnamefont
			{Moya-Cessa}}, \ and\ \bibinfo {author} {\bibfnamefont {D.~N.}\ \bibnamefont
			{Christodoulides}},\ }\bibfield  {title} {\enquote {\bibinfo {title}
			{{E}rmakov-{L}ewis symmetry in photonic lattices},}\ }\href {\doibase
		10.1364/OL.39.002083} {\bibfield  {journal} {\bibinfo  {journal} {Opt.
				Lett.}\ }\textbf {\bibinfo {volume} {39}},\ \bibinfo {pages} {2083--2085}
		(\bibinfo {year} {2014}{\natexlab{a}})}\BibitemShut {NoStop}%
	\bibitem [{\citenamefont {Rodr\'iguez-Lara}\ \emph
		{et~al.}(2014{\natexlab{b}})\citenamefont {Rodr\'iguez-Lara}, \citenamefont
		{Moya-Cessa},\ and\ \citenamefont {Christodoulides}}]{RodriguezLara2014b}%
	\BibitemOpen
	\bibfield  {author} {\bibinfo {author} {\bibfnamefont {B.~M.}\ \bibnamefont
			{Rodr\'iguez-Lara}}, \bibinfo {author} {\bibfnamefont {H.~M.}\ \bibnamefont
			{Moya-Cessa}}, \ and\ \bibinfo {author} {\bibfnamefont {D.~N.}\ \bibnamefont
			{Christodoulides}},\ }\bibfield  {title} {\enquote {\bibinfo {title}
			{Propagation and perfect transmission in three-waveguide axially varying
				couplers},}\ }\href {\doibase 10.1103/PhysRevA.89.013802} {\bibfield
		{journal} {\bibinfo  {journal} {Phys. Rev. A}\ }\textbf {\bibinfo {volume}
			{89}},\ \bibinfo {pages} {013802} (\bibinfo {year} {2014}{\natexlab{b}})},\
	\Eprint {http://arxiv.org/abs/1310.4754} {arXiv:1310.4754 [physics.optics]}
	\BibitemShut {NoStop}%
	\bibitem [{\citenamefont {Rodr\'iguez-Walton}\ \emph
		{et~al.}(2020)\citenamefont {Rodr\'iguez-Walton}, \citenamefont {{Jaramillo
				\'Avila}},\ and\ \citenamefont {Rodr\'iguez-Lara}}]{RodriguezWalton2019}%
	\BibitemOpen
	\bibfield  {author} {\bibinfo {author} {\bibfnamefont {S.}~\bibnamefont
			{Rodr\'iguez-Walton}}, \bibinfo {author} {\bibfnamefont {B.}~\bibnamefont
			{{Jaramillo \'Avila}}}, \ and\ \bibinfo {author} {\bibfnamefont {B.~M.}\
			\bibnamefont {Rodr\'iguez-Lara}},\ }\bibfield  {title} {\enquote {\bibinfo
			{title} {Optical non-{H}ermitian para-{F}ermi oscillators},}\ }\href
	{\doibase 10.1103/PhysRevA.101.043840} {\bibfield  {journal} {\bibinfo
			{journal} {Phys. Rev. A}\ }\textbf {\bibinfo {volume} {101}},\ \bibinfo
		{pages} {043840} (\bibinfo {year} {2020})},\ \Eprint
	{http://arxiv.org/abs/1911.11044} {arXiv:1911.11044 [physics.optics]}
	\BibitemShut {NoStop}%
	\bibitem [{\citenamefont {Marcuse}(1987)}]{Marcuse1987}%
	\BibitemOpen
	\bibfield  {author} {\bibinfo {author} {\bibfnamefont {D.}~\bibnamefont
			{Marcuse}},\ }\bibfield  {title} {\enquote {\bibinfo {title} {Directional
				couplers made of nonidentical asymmetric slabs. {P}art {II}: Grating-assisted
				couplers},}\ }\href {\doibase 10.1109/JLT.1987.1075493} {\bibfield  {journal}
		{\bibinfo  {journal} {J. Lightwave Technol.}\ }\textbf {\bibinfo {volume}
			{5}},\ \bibinfo {pages} {268--273} (\bibinfo {year} {1987})}\BibitemShut
	{NoStop}%
	\bibitem [{\citenamefont {Huang}\ and\ \citenamefont {Haus}(1989)}]{Huang1989}%
	\BibitemOpen
	\bibfield  {author} {\bibinfo {author} {\bibfnamefont {W.}~\bibnamefont
			{Huang}}\ and\ \bibinfo {author} {\bibfnamefont {H.~A.}\ \bibnamefont
			{Haus}},\ }\bibfield  {title} {\enquote {\bibinfo {title} {Power exchange in
				grating-assisted couplers},}\ }\href {\doibase 10.1109/50.32359} {\bibfield
		{journal} {\bibinfo  {journal} {J. Lightwave Technol.}\ }\textbf {\bibinfo
			{volume} {7}},\ \bibinfo {pages} {920--924} (\bibinfo {year}
		{1989})}\BibitemShut {NoStop}%
	\bibitem [{\citenamefont {Griffel}\ and\ \citenamefont
		{Yariv}(1991)}]{Griffel1991}%
	\BibitemOpen
	\bibfield  {author} {\bibinfo {author} {\bibfnamefont {G.}~\bibnamefont
			{Griffel}}\ and\ \bibinfo {author} {\bibfnamefont {A.}~\bibnamefont
			{Yariv}},\ }\bibfield  {title} {\enquote {\bibinfo {title} {Frequency
				response and tunability of grating-assisted directional couplers},}\ }\href
	{\doibase 10.1109/3.83362} {\bibfield  {journal} {\bibinfo  {journal} {IEEE
				J. Quantum Electron.}\ }\textbf {\bibinfo {volume} {27}},\ \bibinfo {pages}
		{1115--1118} (\bibinfo {year} {1991})}\BibitemShut {NoStop}%
	\bibitem [{\citenamefont {Alferness}\ \emph {et~al.}(1992)\citenamefont
		{Alferness}, \citenamefont {Buhl}, \citenamefont {Koren}, \citenamefont
		{Miller}, \citenamefont {Young}, \citenamefont {Koch}, \citenamefont
		{Burrus},\ and\ \citenamefont {Raybon}}]{Alferness1992}%
	\BibitemOpen
	\bibfield  {author} {\bibinfo {author} {\bibfnamefont {R.~C.}\ \bibnamefont
			{Alferness}}, \bibinfo {author} {\bibfnamefont {L.~L.}\ \bibnamefont {Buhl}},
		\bibinfo {author} {\bibfnamefont {U.}~\bibnamefont {Koren}}, \bibinfo
		{author} {\bibfnamefont {B.~I.}\ \bibnamefont {Miller}}, \bibinfo {author}
		{\bibfnamefont {M.~G.}\ \bibnamefont {Young}}, \bibinfo {author}
		{\bibfnamefont {T.~L.}\ \bibnamefont {Koch}}, \bibinfo {author}
		{\bibfnamefont {C.~A.}\ \bibnamefont {Burrus}}, \ and\ \bibinfo {author}
		{\bibfnamefont {G.}~\bibnamefont {Raybon}},\ }\bibfield  {title} {\enquote
		{\bibinfo {title} {Broadly tunable {InGaAsP/InP} buried rib waveguide
				vertical coupler filter},}\ }\href {\doibase 10.1063/1.106480} {\bibfield
		{journal} {\bibinfo  {journal} {Appl. Phys. Lett.}\ }\textbf {\bibinfo
			{volume} {60}},\ \bibinfo {pages} {980--982} (\bibinfo {year}
		{1992})}\BibitemShut {NoStop}%
	\bibitem [{\citenamefont {Weisen}\ \emph {et~al.}(2019)\citenamefont {Weisen},
		\citenamefont {Posner}, \citenamefont {Gates}, \citenamefont {Gawith},
		\citenamefont {Smith},\ and\ \citenamefont {Horak}}]{Weisen2019}%
	\BibitemOpen
	\bibfield  {author} {\bibinfo {author} {\bibfnamefont {M.~J.}\ \bibnamefont
			{Weisen}}, \bibinfo {author} {\bibfnamefont {M.~T.}\ \bibnamefont {Posner}},
		\bibinfo {author} {\bibfnamefont {J.~C.}\ \bibnamefont {Gates}}, \bibinfo
		{author} {\bibfnamefont {C.~B.~E.}\ \bibnamefont {Gawith}}, \bibinfo {author}
		{\bibfnamefont {P.~G.~R.}\ \bibnamefont {Smith}}, \ and\ \bibinfo {author}
		{\bibfnamefont {P.}~\bibnamefont {Horak}},\ }\bibfield  {title} {\enquote
		{\bibinfo {title} {Low-loss wavelength-selective integrated waveguide coupler
				based on tilted {B}ragg gratings},}\ }\href {\doibase
		10.1364/JOSAB.36.001783} {\bibfield  {journal} {\bibinfo  {journal} {J. Opt.
				Soc. Am. B}\ }\textbf {\bibinfo {volume} {36}},\ \bibinfo {pages}
		{1783--1791} (\bibinfo {year} {2019})}\BibitemShut {NoStop}%
	\bibitem [{\citenamefont {Kogelnik}\ and\ \citenamefont
		{Schmidt}(1975)}]{Kogelnik1975}%
	\BibitemOpen
	\bibfield  {author} {\bibinfo {author} {\bibfnamefont {H.~W.}\ \bibnamefont
			{Kogelnik}}\ and\ \bibinfo {author} {\bibfnamefont {R.~V.}\ \bibnamefont
			{Schmidt}},\ }\href@noop {} {}\bibinfo {howpublished} {U.S. Patent No.
		4012113A} (\bibinfo {year} {1975})\BibitemShut {NoStop}%
	\bibitem [{\citenamefont {Kogelnik}\ and\ \citenamefont
		{Schmidt}(1976)}]{Kogelnik1976}%
	\BibitemOpen
	\bibfield  {author} {\bibinfo {author} {\bibfnamefont {H.}~\bibnamefont
			{Kogelnik}}\ and\ \bibinfo {author} {\bibfnamefont {R.}~\bibnamefont
			{Schmidt}},\ }\bibfield  {title} {\enquote {\bibinfo {title} {Switched
				directional couplers with alternating ${\Delta}$${\beta}$},}\ }\href
	{\doibase 10.1109/JQE.1976.1069190} {\bibfield  {journal} {\bibinfo
			{journal} {IEEE J. Quantum Elect.}\ }\textbf {\bibinfo {volume} {12}},\
		\bibinfo {pages} {396--401} (\bibinfo {year} {1976})}\BibitemShut {NoStop}%
	\bibitem [{\citenamefont {Schmidt}\ and\ \citenamefont
		{Alferness}(1979)}]{Schmidt1979}%
	\BibitemOpen
	\bibfield  {author} {\bibinfo {author} {\bibfnamefont {R.}~\bibnamefont
			{Schmidt}}\ and\ \bibinfo {author} {\bibfnamefont {R.}~\bibnamefont
			{Alferness}},\ }\bibfield  {title} {\enquote {\bibinfo {title} {Directional
				coupler switches, modulators, and filters using alternating
				${\Delta}$${\beta}$ techniques},}\ }\href {\doibase 10.1109/TCS.1979.1084592}
	{\bibfield  {journal} {\bibinfo  {journal} {IEEE T. Circuits Syst.}\ }\textbf
		{\bibinfo {volume} {26}},\ \bibinfo {pages} {1099--1108} (\bibinfo {year}
		{1979})}\BibitemShut {NoStop}%
	\bibitem [{\citenamefont {Korotky}(1986)}]{Korotky1986}%
	\BibitemOpen
	\bibfield  {author} {\bibinfo {author} {\bibfnamefont {S.}~\bibnamefont
			{Korotky}},\ }\bibfield  {title} {\enquote {\bibinfo {title} {Three-space
				representation of phase-mismatch switching in coupled two-state optical
				systems},}\ }\href {\doibase 10.1109/JQE.1986.1073030} {\bibfield  {journal}
		{\bibinfo  {journal} {IEEE J. Quantum Elect.}\ }\textbf {\bibinfo {volume}
			{22}},\ \bibinfo {pages} {952--958} (\bibinfo {year} {1986})}\BibitemShut
	{NoStop}%
	\bibitem [{\citenamefont {Findakly}\ and\ \citenamefont
		{Leonberger}(1988)}]{Findakly1988}%
	\BibitemOpen
	\bibfield  {author} {\bibinfo {author} {\bibfnamefont {T.~K.}\ \bibnamefont
			{Findakly}}\ and\ \bibinfo {author} {\bibfnamefont {F.~J.}\ \bibnamefont
			{Leonberger}},\ }\bibfield  {title} {\enquote {\bibinfo {title} {On the
				crosstalk of reversed delta beta directional coupler switches},}\ }\href
	{\doibase 10.1109/50.3960} {\bibfield  {journal} {\bibinfo  {journal} {J.
				Lightwave Technol.}\ }\textbf {\bibinfo {volume} {6}},\ \bibinfo {pages}
		{36--40} (\bibinfo {year} {1988})}\BibitemShut {NoStop}%
	\bibitem [{\citenamefont {Mandel}\ and\ \citenamefont
		{Wolf}(1995)}]{Mandel1995}%
	\BibitemOpen
	\bibfield  {author} {\bibinfo {author} {\bibfnamefont {L.}~\bibnamefont
			{Mandel}}\ and\ \bibinfo {author} {\bibfnamefont {E.}~\bibnamefont {Wolf}},\
	}\href@noop {} {\emph {\bibinfo {title} {Optical Coherence and Quantum
				Optics}}}\ (\bibinfo  {publisher} {Cambridge University Press},\ \bibinfo
	{year} {1995})\BibitemShut {NoStop}%
	\bibitem [{\citenamefont {Turner}(1966)}]{Turner1966}%
	\BibitemOpen
	\bibfield  {author} {\bibinfo {author} {\bibfnamefont {E.~H.}\ \bibnamefont
			{Turner}},\ }\bibfield  {title} {\enquote {\bibinfo {title} {High‐frequency
				electro-optic coefficients of lithium niobate},}\ }\href {\doibase
		10.1063/1.1754449} {\bibfield  {journal} {\bibinfo  {journal} {Appl. Phys.
				Lett.}\ }\textbf {\bibinfo {volume} {8}},\ \bibinfo {pages} {303--304}
		(\bibinfo {year} {1966})}\BibitemShut {NoStop}%
	\bibitem [{\citenamefont {Long}\ \emph {et~al.}(1994)\citenamefont {Long},
		\citenamefont {Myers},\ and\ \citenamefont {Brueck}}]{Long1994}%
	\BibitemOpen
	\bibfield  {author} {\bibinfo {author} {\bibfnamefont {X-C.}\ \bibnamefont
			{Long}}, \bibinfo {author} {\bibfnamefont {R.~A.}\ \bibnamefont {Myers}}, \
		and\ \bibinfo {author} {\bibfnamefont {S.~R.~J.}\ \bibnamefont {Brueck}},\
	}\bibfield  {title} {\enquote {\bibinfo {title} {Measurement of the linear
				electro-optic coefficient in poled amorphous silica},}\ }\href {\doibase
		10.1364/OL.19.001819} {\bibfield  {journal} {\bibinfo  {journal} {Opt.
				Lett.}\ }\textbf {\bibinfo {volume} {19}},\ \bibinfo {pages} {1819--1821}
		(\bibinfo {year} {1994})}\BibitemShut {NoStop}%
	\bibitem [{\citenamefont {Gao}\ \emph {et~al.}(2018)\citenamefont {Gao},
		\citenamefont {Jiang}, \citenamefont {Cui}, \citenamefont {Zhang},
		\citenamefont {Jia},\ and\ \citenamefont {Jiang}}]{Gao2018}%
	\BibitemOpen
	\bibfield  {author} {\bibinfo {author} {\bibfnamefont {H.}~\bibnamefont
			{Gao}}, \bibinfo {author} {\bibfnamefont {Y.}~\bibnamefont {Jiang}}, \bibinfo
		{author} {\bibfnamefont {Y.}~\bibnamefont {Cui}}, \bibinfo {author}
		{\bibfnamefont {L.}~\bibnamefont {Zhang}}, \bibinfo {author} {\bibfnamefont
			{J.}~\bibnamefont {Jia}}, \ and\ \bibinfo {author} {\bibfnamefont
			{L.}~\bibnamefont {Jiang}},\ }\bibfield  {title} {\enquote {\bibinfo {title}
			{Investigation on the thermo-optic coefficient of silica fiber within a wide
				temperature range},}\ }\href {\doibase 10.1109/JLT.2018.2875941} {\bibfield
		{journal} {\bibinfo  {journal} {J. Lightwave Technol.}\ }\textbf {\bibinfo
			{volume} {36}},\ \bibinfo {pages} {5881--5886} (\bibinfo {year}
		{2018})}\BibitemShut {NoStop}%
	\bibitem [{\citenamefont {Liu}\ \emph {et~al.}(2015)\citenamefont {Liu},
		\citenamefont {Ye}, \citenamefont {Khan},\ and\ \citenamefont
		{Sorger}}]{Liu2015}%
	\BibitemOpen
	\bibfield  {author} {\bibinfo {author} {\bibfnamefont {K.}~\bibnamefont
			{Liu}}, \bibinfo {author} {\bibfnamefont {C.~R.}\ \bibnamefont {Ye}},
		\bibinfo {author} {\bibfnamefont {S.}~\bibnamefont {Khan}}, \ and\ \bibinfo
		{author} {\bibfnamefont {V.~J.}\ \bibnamefont {Sorger}},\ }\bibfield  {title}
	{\enquote {\bibinfo {title} {Review and perspective on ultrafast
				wavelength-size electro-optic modulators},}\ }\href {\doibase
		10.1002/lpor.201400219} {\bibfield  {journal} {\bibinfo  {journal} {Laser
				Photonics Rev.}\ }\textbf {\bibinfo {volume} {9}},\ \bibinfo {pages}
		{172--194} (\bibinfo {year} {2015})}\BibitemShut {NoStop}%
	\bibitem [{\citenamefont {Liu}\ \emph {et~al.}(2005)\citenamefont {Liu},
		\citenamefont {Chang},\ and\ \citenamefont {Craig}}]{Liu2005}%
	\BibitemOpen
	\bibfield  {author} {\bibinfo {author} {\bibfnamefont {Y.}~\bibnamefont
			{Liu}}, \bibinfo {author} {\bibfnamefont {T.}~\bibnamefont {Chang}}, \ and\
		\bibinfo {author} {\bibfnamefont {A.~E.}\ \bibnamefont {Craig}},\ }\bibfield
	{title} {\enquote {\bibinfo {title} {Coupled mode theory for modeling
				microring resonators},}\ }\href {\doibase 10.1117/1.2012503} {\bibfield
		{journal} {\bibinfo  {journal} {Opt. Eng.}\ }\textbf {\bibinfo {volume}
			{44}},\ \bibinfo {pages} {1--6} (\bibinfo {year} {2005})}\BibitemShut
	{NoStop}%
	\bibitem [{\citenamefont {Preu}\ \emph {et~al.}(2008)\citenamefont {Preu},
		\citenamefont {Schwefel}, \citenamefont {Malzer}, \citenamefont {D\"ohler},
		\citenamefont {Wang}, \citenamefont {Hanson}, \citenamefont {Zimmerman},\
		and\ \citenamefont {Gossard}}]{Preu2008}%
	\BibitemOpen
	\bibfield  {author} {\bibinfo {author} {\bibfnamefont {S.}~\bibnamefont
			{Preu}}, \bibinfo {author} {\bibfnamefont {H.~G.~L.}\ \bibnamefont
			{Schwefel}}, \bibinfo {author} {\bibfnamefont {S.}~\bibnamefont {Malzer}},
		\bibinfo {author} {\bibfnamefont {G.~H.}\ \bibnamefont {D\"ohler}}, \bibinfo
		{author} {\bibfnamefont {L.~J.}\ \bibnamefont {Wang}}, \bibinfo {author}
		{\bibfnamefont {M.}~\bibnamefont {Hanson}}, \bibinfo {author} {\bibfnamefont
			{J.~D.}\ \bibnamefont {Zimmerman}}, \ and\ \bibinfo {author} {\bibfnamefont
			{A.~C.}\ \bibnamefont {Gossard}},\ }\bibfield  {title} {\enquote {\bibinfo
			{title} {Coupled whispering gallery mode resonators in the {T}erahertz
				frequency range},}\ }\href {\doibase 10.1364/OE.16.007336} {\bibfield
		{journal} {\bibinfo  {journal} {Opt. Express}\ }\textbf {\bibinfo {volume}
			{16}},\ \bibinfo {pages} {7336--7343} (\bibinfo {year} {2008})}\BibitemShut
	{NoStop}%
	\bibitem [{\citenamefont {Haus}\ and\ \citenamefont {Huang}(1991)}]{Haus1991}%
	\BibitemOpen
	\bibfield  {author} {\bibinfo {author} {\bibfnamefont {H.~A.}\ \bibnamefont
			{Haus}}\ and\ \bibinfo {author} {\bibfnamefont {W.}~\bibnamefont {Huang}},\
	}\bibfield  {title} {\enquote {\bibinfo {title} {Coupled-mode theory},}\
	}\href {\doibase 10.1109/5.104225} {\bibfield  {journal} {\bibinfo  {journal}
			{Proc. IEEE}\ }\textbf {\bibinfo {volume} {79}},\ \bibinfo {pages}
		{1505--1518} (\bibinfo {year} {1991})}\BibitemShut {NoStop}%
	\bibitem [{\citenamefont {Elnaggar}\ \emph {et~al.}(2014)\citenamefont
		{Elnaggar}, \citenamefont {Tervo},\ and\ \citenamefont
		{Mattar}}]{Elnaggar2014}%
	\BibitemOpen
	\bibfield  {author} {\bibinfo {author} {\bibfnamefont {S.~Y.}\ \bibnamefont
			{Elnaggar}}, \bibinfo {author} {\bibfnamefont {R.}~\bibnamefont {Tervo}}, \
		and\ \bibinfo {author} {\bibfnamefont {S.~M.}\ \bibnamefont {Mattar}},\
	}\bibfield  {title} {\enquote {\bibinfo {title} {Coupled modes, frequencies
				and fields of a dielectric resonator and a cavity using coupled mode
				theory},}\ }\href {\doibase 10.1016/j.jmr.2013.10.016} {\bibfield  {journal}
		{\bibinfo  {journal} {J. Magn. Reson.}\ }\textbf {\bibinfo {volume} {238}},\
		\bibinfo {pages} {1--7} (\bibinfo {year} {2014})}\BibitemShut {NoStop}%
	\bibitem [{\citenamefont {Elnaggar}\ \emph {et~al.}(2015)\citenamefont
		{Elnaggar}, \citenamefont {Tervo},\ and\ \citenamefont
		{Mattar}}]{Elnaggar2015}%
	\BibitemOpen
	\bibfield  {author} {\bibinfo {author} {\bibfnamefont {S.~Y.}\ \bibnamefont
			{Elnaggar}}, \bibinfo {author} {\bibfnamefont {R.~J.}\ \bibnamefont {Tervo}},
		\ and\ \bibinfo {author} {\bibfnamefont {S.~M.}\ \bibnamefont {Mattar}},\
	}\bibfield  {title} {\enquote {\bibinfo {title} {Energy coupled mode theory
				for electromagnetic resonators},}\ }\href {\doibase
		10.1109/TMTT.2015.2434377} {\bibfield  {journal} {\bibinfo  {journal} {IEEE
				T. Microw. Theory}\ }\textbf {\bibinfo {volume} {63}},\ \bibinfo {pages}
		{2115--2123} (\bibinfo {year} {2015})},\ \Eprint
	{http://arxiv.org/abs/1305.6085} {arXiv:1305.6085 [cond-mat.other]}
	\BibitemShut {NoStop}%
	\bibitem [{\citenamefont {Kim}\ and\ \citenamefont {Ling}(2007)}]{Kim2007}%
	\BibitemOpen
	\bibfield  {author} {\bibinfo {author} {\bibfnamefont {Y.}~\bibnamefont
			{Kim}}\ and\ \bibinfo {author} {\bibfnamefont {H.}~\bibnamefont {Ling}},\
	}\bibfield  {title} {\enquote {\bibinfo {title} {Investigation of coupled
				mode behaviour of electrically small meander antennas},}\ }\href {\doibase
		10.1049/el20072165} {\bibfield  {journal} {\bibinfo  {journal} {Electron.
				Lett.}\ }\textbf {\bibinfo {volume} {43}},\ \bibinfo {pages} {1250--1252}
		(\bibinfo {year} {2007})}\BibitemShut {NoStop}%
	\bibitem [{\citenamefont {Agarwal}\ \emph {et~al.}(2006)\citenamefont
		{Agarwal}, \citenamefont {Sylvester},\ and\ \citenamefont
		{Blaauw}}]{Agarwal2006}%
	\BibitemOpen
	\bibfield  {author} {\bibinfo {author} {\bibfnamefont {K.}~\bibnamefont
			{Agarwal}}, \bibinfo {author} {\bibfnamefont {D.}~\bibnamefont {Sylvester}},
		\ and\ \bibinfo {author} {\bibfnamefont {D.}~\bibnamefont {Blaauw}},\
	}\bibfield  {title} {\enquote {\bibinfo {title} {Modeling and analysis of
				crosstalk noise in coupled {RLC} interconnects},}\ }\href {\doibase
		10.1109/TCAD.2005.855961} {\bibfield  {journal} {\bibinfo  {journal} {IEEE T.
				Comput. Aid. D.}\ }\textbf {\bibinfo {volume} {25}},\ \bibinfo {pages}
		{892--901} (\bibinfo {year} {2006})}\BibitemShut {NoStop}%
	\bibitem [{\citenamefont {Snyder}(1972)}]{Snyder1972}%
	\BibitemOpen
	\bibfield  {author} {\bibinfo {author} {\bibfnamefont {A.~W.}\ \bibnamefont
			{Snyder}},\ }\bibfield  {title} {\enquote {\bibinfo {title} {Coupled-mode
				theory for optical fibers},}\ }\href {\doibase 10.1364/JOSA.62.001267}
	{\bibfield  {journal} {\bibinfo  {journal} {J. Opt. Soc. Am.}\ }\textbf
		{\bibinfo {volume} {62}},\ \bibinfo {pages} {1267--1277} (\bibinfo {year}
		{1972})}\BibitemShut {NoStop}%
	\bibitem [{\citenamefont {McIntyre}\ and\ \citenamefont
		{Snyder}(1973)}]{McIntyre1973}%
	\BibitemOpen
	\bibfield  {author} {\bibinfo {author} {\bibfnamefont {P.~D.}\ \bibnamefont
			{McIntyre}}\ and\ \bibinfo {author} {\bibfnamefont {A.~W.}\ \bibnamefont
			{Snyder}},\ }\bibfield  {title} {\enquote {\bibinfo {title} {Power transfer
				between optical fibers},}\ }\href {\doibase 10.1364/josa.63.001518}
	{\bibfield  {journal} {\bibinfo  {journal} {J. Opt. Soc. Am.}\ }\textbf
		{\bibinfo {volume} {63}},\ \bibinfo {pages} {1518--1527} (\bibinfo {year}
		{1973})}\BibitemShut {NoStop}%
	\bibitem [{\citenamefont {Huang}(1994)}]{Huang1994}%
	\BibitemOpen
	\bibfield  {author} {\bibinfo {author} {\bibfnamefont {W.-P.}\ \bibnamefont
			{Huang}},\ }\bibfield  {title} {\enquote {\bibinfo {title} {Coupled-mode
				theory for optical waveguides: an overview},}\ }\href {\doibase
		10.1364/JOSAA.11.000963} {\bibfield  {journal} {\bibinfo  {journal} {J. Opt.
				Soc. Am. A}\ }\textbf {\bibinfo {volume} {11}},\ \bibinfo {pages} {963--983}
		(\bibinfo {year} {1994})}\BibitemShut {NoStop}%
	\bibitem [{\citenamefont {Wei}\ and\ \citenamefont {Norman}(1963)}]{Wei1963}%
	\BibitemOpen
	\bibfield  {author} {\bibinfo {author} {\bibfnamefont {J.}~\bibnamefont
			{Wei}}\ and\ \bibinfo {author} {\bibfnamefont {E.}~\bibnamefont {Norman}},\
	}\bibfield  {title} {\enquote {\bibinfo {title} {{L}ie algebraic solution of
				linear differential equations},}\ }\href {\doibase 10.1063/1.1703993}
	{\bibfield  {journal} {\bibinfo  {journal} {J. Math. Phys.}\ }\textbf
		{\bibinfo {volume} {4}},\ \bibinfo {pages} {575--581} (\bibinfo {year}
		{1963})}\BibitemShut {NoStop}%
	\bibitem [{\citenamefont {Allen}\ and\ \citenamefont
		{Eberly}(1975)}]{Allen1975}%
	\BibitemOpen
	\bibfield  {author} {\bibinfo {author} {\bibfnamefont {L.}~\bibnamefont
			{Allen}}\ and\ \bibinfo {author} {\bibfnamefont {J.~H.}\ \bibnamefont
			{Eberly}},\ }\href@noop {} {\emph {\bibinfo {title} {Optical resonance and
				two-level atoms}}}\ (\bibinfo  {publisher} {Wiley},\ \bibinfo {year}
	{1975})\BibitemShut {NoStop}%
	\bibitem [{\citenamefont {Eaton}\ \emph {et~al.}(2011)\citenamefont {Eaton},
		\citenamefont {Ng}, \citenamefont {Osellame},\ and\ \citenamefont
		{Herman}}]{Eaton2011}%
	\BibitemOpen
	\bibfield  {author} {\bibinfo {author} {\bibfnamefont {S.~M.}\ \bibnamefont
			{Eaton}}, \bibinfo {author} {\bibfnamefont {M.~L.}\ \bibnamefont {Ng}},
		\bibinfo {author} {\bibfnamefont {R.}~\bibnamefont {Osellame}}, \ and\
		\bibinfo {author} {\bibfnamefont {P.~R.}\ \bibnamefont {Herman}},\ }\bibfield
	{title} {\enquote {\bibinfo {title} {High refractive index contrast in fused
				silica waveguides by tightly focused, high-repetition rate femtosecond
				laser},}\ }\href {\doibase 10.1016/j.jnoncrysol.2010.11.082} {\bibfield
		{journal} {\bibinfo  {journal} {J. Non-Cryst. Solids}\ }\textbf {\bibinfo
			{volume} {357}},\ \bibinfo {pages} {2387--2391} (\bibinfo {year}
		{2011})}\BibitemShut {NoStop}%
	\bibitem [{\citenamefont {Tsukada}(1977)}]{Tsukada1977}%
	\BibitemOpen
	\bibfield  {author} {\bibinfo {author} {\bibfnamefont {N.}~\bibnamefont
			{Tsukada}},\ }\bibfield  {title} {\enquote {\bibinfo {title} {Modification of
				the coupling coefficient by periodic modulation of the propagation
				constants},}\ }\href {\doibase 10.1016/0030-4018(77)90259-0} {\bibfield
		{journal} {\bibinfo  {journal} {Opt. Commun.}\ }\textbf {\bibinfo {volume}
			{22}},\ \bibinfo {pages} {113--115} (\bibinfo {year} {1977})}\BibitemShut
	{NoStop}%
	\bibitem [{\citenamefont {Molter-Orr}\ and\ \citenamefont
		{Haus}(1985)}]{Molter1985}%
	\BibitemOpen
	\bibfield  {author} {\bibinfo {author} {\bibfnamefont {L.~A.}\ \bibnamefont
			{Molter-Orr}}\ and\ \bibinfo {author} {\bibfnamefont {H.~A.}\ \bibnamefont
			{Haus}},\ }\bibfield  {title} {\enquote {\bibinfo {title} {Multiple coupled
				waveguide switches using alternating ${\Delta}$${\beta}$ phase mismatch},}\
	}\href {\doibase 10.1364/AO.24.001260} {\bibfield  {journal} {\bibinfo
			{journal} {Appl. Opt.}\ }\textbf {\bibinfo {volume} {24}},\ \bibinfo {pages}
		{1260--1264} (\bibinfo {year} {1985})}\BibitemShut {NoStop}%
\end{thebibliography}
%%%%%%%%%%%%%%%%%%%%%%% References %%%%%%%%%%%%%%%%%%%%%%%%%
%%%%%%%%%%%%%%%%%%%%%%% References %%%%%%%%%%%%%%%%%%%%%%%%%	
\end{document}